\documentclass[useAMS,usenatbib]{mn2e}

\usepackage{natbib}
\usepackage{amsmath}
\usepackage{amsfonts}
\usepackage{amssymb}
\usepackage{amsbsy} 
\usepackage{graphicx}
\usepackage{setspace}
\usepackage{deluxetable}
\newcommand{\aj}{Astronomical Journal}
\newcommand{\araa}{Ann. Rev. Ast. \& Ast.}
\newcommand{\apj}{ApJ}
\newcommand{\apjl}{ApJL}
\newcommand{\apjs}{ApJS}
\newcommand{\ao}{Applied Optics}

\newcommand{\aap}{Astronomy and Astrophysics}

\newcommand{\mnras}{MNRAS}

\newcommand{\pasp}{Publications of the ASP}

\newcommand{\rmxaa}{Revista Mexicana de Astronom\'ia y Astrof\'isica}

\title[Induced nuclear activity in galaxy pairs with different morphologies (E+E), (E+S) and (S+S).]{Induced nuclear activity in galaxy pairs with different morphologies (E+E), (E+S) and (S+S).}
\author[F. J. Hern\'andez-Ibarra et al.]
{Francisco J. Hern\'andez-Ibarra$^{1}$\thanks{E-mail:hibarra@astro.unam.mx}, Yair Krongold$^{1}$, Deborah Dultzin$^{1}$,
\newauthor Ascensi\'on del Olmo$^{2}$, Jaime Perea$^{2}$, Jes\'us Gonz\'alez$^{1}$,
\newauthor Sandro Mendoza-Castrej\'on$^{1}$ and Theodoros Bitsakis$^{1}$ \\
$^{1}$Instituto de Astronom\'ia, Universidad Nacional Aut\'onoma de M\'exico, Apartado Postal 70-264, 04510 M\'exico DF, M\'exico.\\
$^{2}$Instituto de Astrof\'isica de Andaluc\'ia (C.S.I.C.) Apartado 3004, 18080 Granada, Spain.}
\begin{document}

\date{}

 \pagerange{\pageref{firstpage}--\pageref{lastpage}} \pubyear{2002}

\maketitle

 \label{firstpage}

\begin{abstract}
We analysed 385 galactic spectra from the Sloan Digital Sky Survey Data Release 7 (SDSS-DR7) that belong to the catalog of isolated pairs of galaxies by Karachentsev.
The spectra corresponds to physical pairs of galaxies as defined by ÎV $\leq$ 1200 Km/s and a pair separation $\leq$ 100 kpc. We search for the incidence of nuclear activity, both
 thermal (star-forming) and non-thermal -Active Galactic Nuclei (AGN)-.
 After a careful extraction of the nuclear spectra, we use diagnostic diagrams and find that the incidence of AGN activity is 48 \% in the paired galaxies with emission lines and 40\% for the total sample (as compared to $\sim$ 43 \% and 41\% respectively in a sample of isolated galaxies).
 These results remain after dissecting the effects of morphological type and galactic stellar mass (with only a small, non significant, enhancement of the AGN fraction in pairs of objects). These results suggest that weak interactions are not necessary or sufficient to trigger low-luminosity AGN. Since the fraction of AGN is predominant in early type spiral galaxies, we conclude that the role of a bulge, and a large gas reservoir are both essential for the triggering of nuclear activity.
The most striking result is that type 1 galaxies are almost absent from the AGN sample. This result
 is in conflict with the Unified Model, and suggests that high accretion rates are essential to form the Broad Line Region in active galaxies. 
\end{abstract}


\section{Introduction}

One of the outstanding problems in the understanding of the Active Galactic Nuclei
(AGN) phenomenon is the feeding processes of the central Supermassive Black Hole (SMBH). The gas fueling may be driven from extragalactic to galactic,
 and further to nuclear scales. 
 The main proposed mechanism to induce gas inflow to the centre of galaxies, on the extragalactic and galactic scales, consists primarily of interactions with 
other galaxies \citep{Ba92, Hop05, spr05, cox08, elli08, kna09}.  In this paper we address  observational evidence of the role of gravitational interactions 
in inducing nuclear activity.

\subsection{Studying the Environment of AGN}
In the past 20 years several efforts  have focused in the study of the 
environment of AGN, in an attempt to elucidate this question from an observational point of view.  
Most of the investigations have dealt with samples of Seyfert galaxies, because these are the closest clearly non-thermal dominated active nuclei. 
Low Ionization Nuclear Emission Regions (LINERs) are easy to
observe, however, the nature of the dominating emission mechanism is still debated \citep{Kron03, 2007IAUS..238..373G, Gon15}. Starburst phenomena
(particularly circumnuclear) have also been included as a type (and/or part) of activity.  The first authors to propose a Starburst-AGN connection 
were \citet{1985MNRAS.213..665P}. 
 An excellent review on this topic can be found in \citet{2008RMxAC..32..139S}.

%\begin{table*}\centering 
%\caption{General statistics on every single galaxy subsample pair.}  
 %\begin{tabular}{cccccccc}
%\hline
%Sample &  Total & No Emission &  Excluded & Fitted & AGN+Comp	& H {\scriptsize II} & AGN Type 1 \\
%\hline
%(E+E) & 58 & 30(52\%) & 3(5\%) & 25(43\%) & 19(33\%) & 6(10\%) & 2(3\%)\\
%(E+S) & 103 & 23(22\%) & 2(2\%) & 78(76\%) & 48(47\%) & 30(29\%) & 1(1\%)\\
%(S+S) & 224 & 9(4\%) & 9(4\%) & 203(91\%) & 83(37\%) & 120(54\%) & 1(0.4\%)\\
%\hline
%TOTAL & 385 & 63(16\%) & 14(4\%) & 306(79\%) & 150(39\%) & 156(41\%) & 4(1\%)\\
%\hline
%\end{tabular}
%\end{table*}

\begin{figure*}
 \includegraphics[scale=1.2]{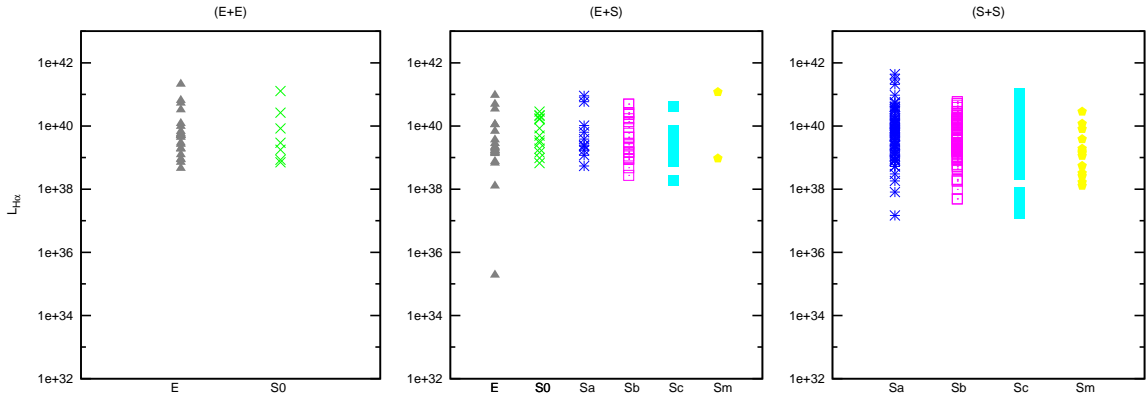}
 \caption{Morphological distribution of H$\alpha$ luminosity for: left;
 (E+E) sample, middle; (E+S) sample and right; (S+S) sample. Mean values of AGN H$\alpha$
 luminosity for the three samples are \emph{L$_H$$_\alpha$ = 2.28
 $\times$ 10$^{40}$ erg s$^{-1}$}, \emph{L$_H$$_\alpha$ = 2.92
 $\times$ 10$^{40}$ erg s$^{-1}$} and  \emph{L$_H$$_\alpha$ = 1.44
 $\times$ 10$^{40}$ erg s$^{-1}$} respectively.}
 \end{figure*}

The first studies of the extragalactic influence on AGN, were devoted to investigate the difference in the
environment between active and non-active galaxies, without distinguishing activity type, \citep{1984PhDT.........5D, 1985ApJS...57..643D}. But soon it became clear that it was
necessary to distinguish between  type 1 and type 2 AGN (and even Starburst or enhanced Star Forming activity).  More recently, the importance of making a difference
between close and large scale environment has become clear. Seminal studies were affected by 
the lack of  clear sample definitions, statistical biases and also biases introduced by sample selection effects.
All these methodological problems  yielded 
contradictory results that can be found in the literature for over more than 20 years; 
from \citet{1982ApJS...50..517S, 1982ApJ...262...66S} to \citet{Kou13}. 
 One of the first discussions of these effects is given in \citet{Dul99} hereafter DH99, and a detailed account of these 
different results is reviewed and analyzed by \citet{2006A&A...451..809S}.

As more refined studies have
been performed, it has become clear that Seyfert 2 galaxies are
in interaction with the same frequency than Star-Forming Galaxies (SFG)
\citep{Kron01, 2008RMxAC..32..139S}, while Seyfert 1 galaxies are in interaction less frequently. Seyfert 1s are found with close companions 
comparably as often as non-active galaxies \citep[DH99]{Kron02}. The most recent
studies confirm these findings considering only physical companions,
i.e. not only from statistical considerations, but from actual
measurements of radial velocities for the neighboring galaxies (Koulouridis et al. 2006a,b).  From a statistical point 
of view,  Sorrentino et al. (2006) have made the comparison of the environment of AGN, SFG and normal
galaxies, for a complete sample of  1829 Seyfert galaxies (725 Sy1 and 1104 Sy2) and 6061 SFG from the Fourth Data Release (DR4) of the SDSS. 
  This study fully confirms the results found by DH99 and by \citet{2006ApJ...651...93K, 2006ApJ...639...37K}. 
 The authors state that for close systems ($\leq$ 100 kpc) they find a
a higher fraction of Sy2 compared to Sy1 in agreement with DH99,
moreover, the frequency of Sy2 is similar to that of SFGs. The most recent confirmation of this result
comes from Villarroel \& Korn (2014). At large scales however, there is no strong evidence of a denser environment for AGN compared to ``normal'' galaxies, in agreement with \citet{2001AJ....122.2243S} and \citet{2006ApJ...651...93K}. Any difference in the large-scale environment of
Sy2 and Sy1 is more related to the morphological  type of the host rather than to activity \citep[see also][]{2008RMxAC..32..150M}.

\subsection{A Complementary Approach: Studying the Incidence of Nuclear Activity in Interacting Galaxies}
In all of the former analysis, the environment of well defined samples of active
vs. non-active galaxies were compared. In the present paper we adopt a
complementary approach. We study the incidence of nuclear activity in
a well defined sample of interacting galaxies. We focus on the
sample of the Catalogue of Isolated Pairs in the Northern Hemisphere \citep{Kara72}. In order to quantify the incidence of 
nuclear activity in the pairs of our sample, we distinguish three morphological pairs; 1) Elliptical plus Elliptical pair (E+E)
 considering galaxies with spheroidal morphology S0, 2) Elliptical plus Spiral pair (E+S) and 3) Spiral plus Spiral pair (S+S) 

There has been efforts in the literature to asses the incidence of AGN in pairs of galaxies. 
Galaxies in elliptical pairs (E+E) have shown an enhancement on the level of recent star formation relative to a control sample of early-type galaxies 
\citep{Ro09}. After a first stage of an encounter that triggers residual star formation, a more efficient inflow of gas towards the 
centre may switch the object to an AGN phase. Thus the possibility that external
perturbations may enhance the frequency of nuclear activity among
galaxies  has been suggested by previous studies \citep{Dul99, Kron02,
Kron03, Ro09, 2011MNRAS.418.2043E, Vil12, liu12} and supported later by mid-IR spectroscopy \citep[e.g.][]{Men15}.

Mixed galaxy pairs (E+S) are a unique laboratory to study the effect of tidal
forces in triggering nuclear activity because they are  
relatively simple systems where a gas rich galaxy interacts with a gas poor one. 
In such systems a clean interpretation of the origin and evolution of the fueling material is possible. 
Mixed pairs minimize the role of the relative
orientation and pair component spin vectors, in driving 
interaction-induced effects \citep{1993AJ....106.1771K}. Since the late-type spiral component is the primary 
source of gas in a mixed pair, it is therefore expected to be the site of all or most star formation 
and nuclear activity, although recent results have shown evidence of
star formation and AGN activity in a non-negligible fraction of the early type components of the
pairs  \citep[based on IRAS data]{1995IAUS..164..434D, 1996A&A...308..387D, 2005AJ....129.2579D}. The presence of
AGN activity on the early type components can be directly confirmed by means of
spectroscopic data \citep{Dul08, 2011RMxAA..47..361C, Sa12}.

Previous studies, based on spiral-spiral pairs (S+S), have shown that
starburst and possibly AGN activity in galaxies may be triggered by interactions.
\citet{1984ApJ...279L...5K} studied a sample of 56 nearby spirals in pairs vs. a control 
sample of 86 non-interacting galaxies, and found that interactions induce an enhancement of the level of 
nuclear activity.  Furthermore, they also found a significant fraction of 
Seyfert or Seyfert-like type nuclei.  However, these studies have searched for activity
without distinguishing between thermal (starburst, hereafter SB) and
non-thermal (properly an AGN) activity. Thus they have not
addressed the incidence of type 1 vs. type 2 AGN in pairs of galaxies.

The incidence of AGN in groups have also been addressed by several authors. 

\citet{2008ApJ...678L...9M, 2010AJ....139.1199M} studied compact
groups of galaxies, consisting of associations of 4 galaxies or more with 
very high local densities in an rather isolated large scale environment. A high frequency of nuclear activity is observed for 270 galaxies in 
64 Hickson compact groups. In particular it is found that among the emission line galaxies
(63\% of the whole sample), 45\% show a pure AGN, 23\% have
a composite spectrum and 32\% show nuclear star formation.
In all cases, the nuclear activity is
manifested as low luminosity AGN and there is a statistically significant deficiency of type 1 AGN as compared
to field galaxies. \citet{Bits15} study the evolution of the nuclear activity and
how it has been affected by the dense environment of the groups. Their analysis is
based on the largest multiwavelength compact group sample to-date.
 They observe that over the past 3 Gyr,  there has been an enhancement of 15\%  in the number of the AGN-hosting late-type
galaxies. This enhancement is accompanied by the corresponding decrease of their circumnuclear star formation. These authors also
show that at any given stellar mass, galaxies found in dynamically
old groups are more likely to host an AGN, than those in young groups.

When going to richer environments, such as clusters of galaxies, the AGN fraction
decreases. However, the results seem to depend on the methods to search for the AGNs. In X-rays,
AGNs are detected with low fraction \citep[$\sim$5-8\%;][]{2007ApJ...664..761M}, with an increasing frecuency
at higher redshift and AGN luminosity \citep {2013ApJ...768....1M}. In the optical
\citet {2006A&A...460L..23P} found a LLAGN fraction of 15 to 21\%, when analysing SDSS spectroscopy of 324 nearby clusters.
They  report a clear increasing trend of the AGN fraction as the cluster
velocity dispersion decreases and the merging rate increases. All these richness of results
clearly indicates that in order to elucidate the exact effect of interactions on triggering AGN, it is mandatory to go to systems such as isolated 
pairs of galaxies, where the effects of single galaxy-galaxy encounters (and the lack of further perturbations) can provide a clear answer.

In this work, we study the incidence of nuclear activity in a well defined sample of galaxy pairs. Host spectra galaxy extraction 
and line emission measurements have been performed systematically in order to make a good quantification of the data. We define the activity 
type of AGN and separate between Seyfert and LINER nuclei. 

In the following section we present the sample and data analysis (\S 2), where an optical classification by diagnostic 
diagrams is performed. On \S 3 we present our results on the incidence of nuclear activity. Discussion takes place on 
\S 4. On section 5 we present the final conclusions of this work.  Throughout the paper, we compare our results on pairs of galaxies, with those of a rigorously defined sample of isolated galaxies (results by our group: Hern\'andez-Ibarra et al. 2013, hereafter H-I13). The methods and analysis between  H-I13 and this work are fully self-consistent, which warrants a direct and reliable comparison. H-I13 analyze to different samples of isolated galaxies. Throughout the paper we compare with the results over the sample defined with the catalogue of isolated galaxies
(CIG;  Karachentseva 1972), for which we obtain more reliable results (as discussed in H-I13).

\section{Sample and data analysis}

This study is based on the Catalog of Isolated Pairs of galaxies \citep[CPG]{Kara72} that contains nearly 600 galaxy pairs. 
The catalogue is based on a visual  search of the Palomar Sky Survey with $\delta$ $\geq$ -3$^{\textrm {o}}$. The majority of the objects 
have high galactic latitude $b$ $\geq$ 20$^{\textrm {o}}$ (in order to avoid galactic extinction) and magnitude limit m$_{Zw}$ $\leq$ 15.7. \citet{Kara72}
used a strong pair-isolation criterion in terms of the apparent angular separation between pairs ($\leq$ 100 Kpc).
The criteria used to build the CPG can be resumed by the following relations:

\begin{equation}
  \frac{x_{ij}}{x_{12}} \geq \frac{5a_i}{a_j};\hspace{0.5cm}0.5a_j\leq a_i\leq4a_j    
\end{equation}\\
where $j=1,2$ correspond to the pair components and $i$ is the nearest neighbour, $a$ represents the major-axis diameter and $x$ the apparent
separation. The overall completeness of this catalogue has been estimated in $\sim$ 90 $\%$ \citep{Her-Tol99}.

We obtain all the available spectra for our sample from the Sloan Digital Sky Survey Data Release 7 (SDSS-DR7) \citep{2009ApJS..182..543A}. The spectra have a wavelength coverage from 
3800-9200\AA{} and a resolution of 1800-2200 with a signal-noise \textgreater 4 per pixel at g=20.2.  We find that 99$\%$ of the  objects 
in our sample have a r-magnitud brigther than the SDSS spectroscopic limit (r = 17.77), thus our sample has a high completeness.
We note that in several cases, only one of the member of the pairs have spectroscopic data, and thus, only that object is included in the analysis.
The total sample consists of 385 galaxies for which spectral data is available.  Tables \ref{a1}1,  \ref{a2}2, and \ref{a3}3 present the samples for the different types of pairs.

\begin{figure}
\includegraphics[scale=0.65]{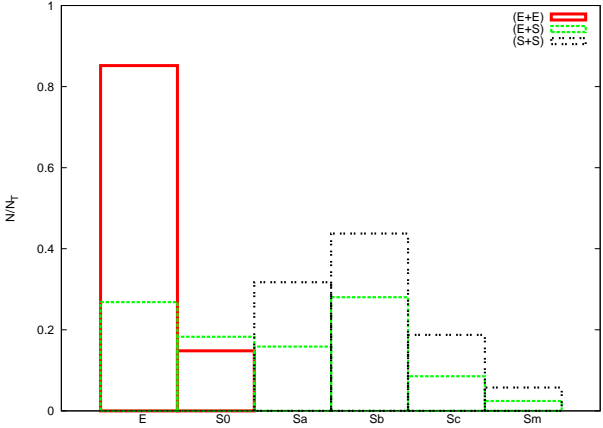}
\caption{Morphological distribution of galaxies in elliptical, mixed and spiral pairs. Continuous red line correspond to (E+E) pair sample, 
green dashed line to (E+S) and black dotted line to (S+S) pair samples.}
\end{figure}

\begin{figure}
\includegraphics[scale=0.65]{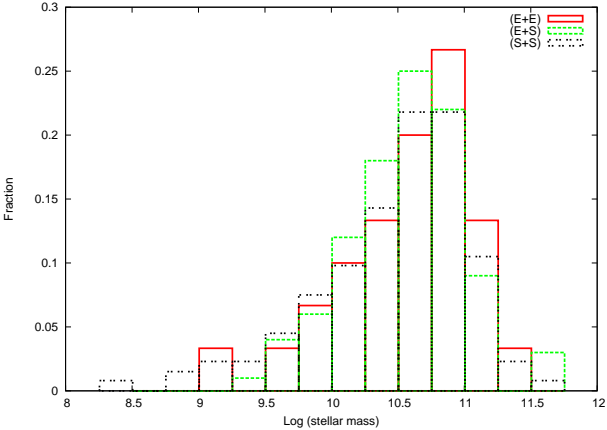}
\caption{Mass distribution for the three subsample galaxy pairs. Labels like in Fig. 2.}
\end{figure}

\begin{figure}
\includegraphics[scale=0.45]{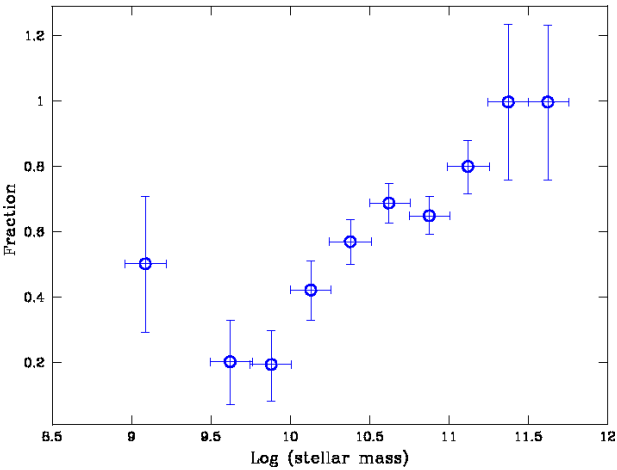}
\caption{AGN fraction of galaxies as a function of stellar mass. AGN are show as blue empty circles.
 Errors in y-direction are the standard deviation per bin and the ``error bars'' in x-direction denote the range of mass in each bin.}
\end{figure}

 We note that the SDSS spectra were taken through a fibre aperture of 3 arcsec in diameter (corresponding to 1.25 kpc at a redshift of 0.02).  This guarantees that the nuclear region of the galaxies was observed. However, the collected integrated spectroscopic area is large enough to include stellar light contamination. This contamination  turns out to be  more significant at the central parts of the galaxies as the spheroidal/bulge component
 becomes more relevant.  Therefore, in order to obtain a 
reliable 
nuclear classification based on the emission lines it is mandatory to subtract the stellar contribution. We applied the principal component analysis 
(PCA) 
method following \citet{2005AJ....129.1795H} to subtract this contribution. We used their first 8 eigenspectra from their low redshift range. These 
eigenspectra are the resulting
 eigenvectors of a PCA analysis applied to a sample of high S/N spectra of non-emission galaxies. In addition, as they pointed out, we included 
two more 
components, an A star spectrum accounting for the possible presence of poststarburst features and a power-law to take into account the possible
 existence of a non-thermal component. The analysis is performed for all the spectra of our sample and it consists on a multiple regression of 
each spectrum to a linear combination of the 8 
eigenspectra plus the two additional components. Previously to the fit each galaxy spectrum was moved to zero redshift which is the one of the 
template library.
 We also masked all those regions where emission lines may appear since the quality of the fit lies on the matching of the continua. Once the 
regression is
 performed, the direct subtraction of the resulting fit to the original (z=0) spectrum provides us with a pure emission line spectrum where all 
the underlying
 absorption components and eventually a non thermal component of the continuum are removed.

With a clean emission line spectrum at hand, we focused in measuring the intensity of the 7 strongest emission lines: H$\beta$, [O {\scriptsize III}] $\lambda$5007,
[O {\scriptsize I}] $\lambda$6300, H$\alpha$, [N {\scriptsize II}] $\lambda$6584 and [S {\scriptsize II}] $\lambda$$\lambda$6717,31. We consider clear line detections those features with a signal to noise ratio (S/N)$\geq$ 3. In our measurements, we were careful to distinguish between intrinsic no detected emission (to a given flux limit) and a
lack of detectability related to low S/N data. For this purpose, we set a
threshold of 10$^{38}$ erg s$^{-1}$ in the H$\alpha$ luminosity.  We consider 
galaxies below this threshold as true non-emission objects (according to our results, the probability that a galaxy with such low H$\alpha$ luminosity is an AGN is less than 2\%, which gives high robustness to our results). Objects with no detected lines, but a 3 $\sigma$ upper limit larger than this threshold in the  H$\alpha$ line are not included in the results, as it is not clear whether these are no-emission line galaxies, or just objects with poor data quality. In this way a systematic and self-consistent luminosity limited analysis is warranted. The distribution of morphology and H$\alpha$ luminosity of our sample (presented for each morphological pair type) are presented in Fig. 1.

From our 385 spectra, 63 are true non-emission objects, 9 do not have all the 7 emission lines detected with a significance above 3 $\sigma$, 4 have problems with the host galaxy spectra subtraction, and 1 does not have the optic fibre in the galaxy centre.  Objects with detections below 3 $\sigma$, without good subtraction and with the fibre off the center were not included in this work (14 objects).

%There are 7 spectra which present less than 3 $\sigma$ detection in their lines. KPG 303B in H$\beta$ and [O {\scriptsize I}] $\lambda$6300 
%lines has $>$ 1  $\sigma$, KPG 317A with [O {\scriptsize I}] $\lambda$6300 $>$ 2 $\sigma$, KPG 339B has H$\beta$ with $>$ 2  $\sigma$ 
%detection and [O {\scriptsize I}] $\lambda$6300 $>$ 1  $\sigma$, KPG 392A with H$\beta$ $<$ 1  $\sigma$ and [O {\scriptsize I}] $\lambda$6300 $>$ 1
  %$\sigma$, KPG 408A has $>$ 2  $\sigma$ and [O {\scriptsize I}] $\lambda$6300 $>$ 1  $\sigma$, KPG 412B with H$\beta$ $>$ 1  $\sigma$,
 %[O {\scriptsize I}] $\lambda$6300 $>$ 2  $\sigma$ and [O {\scriptsize III}] $\lambda$5007 $>$ 2  $\sigma$ and KPG 432A with H$\beta$ $>$ 2  $\sigma$, 
%[O {\scriptsize I}] $\lambda$6300 $<$ 1. KPG 058B does not have a good host galaxy extraction, so, we did not include in the sample. 
%KPG 445A has only [N {\scriptsize II}] $\lambda$6584 and H$\alpha$ lines above 3 $\sigma$ detection but, it can be possible make a preliminary
% classification with only this two lines as it shows on Fig. 6(a). 

 Fluxes were calculated with the Sherpa software
(http://cxc.cfa.harvard.edu/sherpa/) which comes in the CIAO distribution, http://cxc.harvard.edu/ciao/. This program fits emission
 lines with Gaussians.  

We evaluated the line intensities using two methods to fit the lines (see full description in H-I13). The first one consisted
in constraining the width and velocity to the same value in our
fits for two separated groups of lines, forbidden and permitted.
Therefore, these fits included four free parameters in our models:
the width and velocity of the forbidden and the permitted lines. An
additional free parameter for each line in the model was the intensity.
The second method constrains
the width and velocity for all the detected lines (independently of
whether they were permitted or forbidden) to have the same value
(i.e. only two free parameters to model all lines) plus an additional
free parameter for the intensity of each line. Our results show that
both methods are equivalent without any substantial difference.
We report here the emission line intensities obtained with the second method
that has less free parameters. 
We note that for those objects where a broad component was required in addition to the narrow one, an individual broad
Gaussian was fitted with fully independent free parameters.

Figs. 2 and 3 show the morphological type and mass distributions respectively for the galaxies in the three subsamples of galaxy pairs.
In Fig. 4 we show  the AGN and HII activity fraction versus the logarithm of stellar mass per bin, derived for our entire sample. 

%From this figure we can see a tendency for higher mass galaxies to host
% an AGN, while lower mass emission line galaxies tend to be H {\scriptsize II} galaxies.

\subsection{Optical classification}

The optical classification  was performed using diagnostic diagrams \citep{Bal81, 1987ApJS...63..295V}. Using the 7 emission lines measured, diagnostic diagrams were constructed with the aid of line ratios [O {\scriptsize III}]/H$\beta$, [N {\scriptsize II}]/H$\alpha$, [S {\scriptsize II}]/H$\alpha$ and [O {\scriptsize I}]/H$\alpha$. 

We based our activity type classification on the  ([N {\scriptsize II}]/H$\alpha$) versus ([O {\scriptsize III}]/H$\beta$)  diagnostic diagram (hereafter [N {\scriptsize II}] diagram), which is the main diagnostic to distinguish objects with different nature. However, we have used the ([S {\scriptsize II}]/H$\alpha$) versus ([O {\scriptsize III}]/H$\beta$) diagnostic diagram (hereafter [S {\scriptsize II}] diagram) to  separate between star-forming galaxies, Seyfert galaxies and LINERs. We have chose to do this, rather than using the  Schawinski et al. (2007) diagnostic over the [N {\scriptsize II}] diagram in order to keep the results of this work consistent with those by H-I13. We note, nevertheless, that the [N {\scriptsize II}] and [S {\scriptsize II}] diagrams give consistent results. We do not use the the ([O {\scriptsize I}]/H$\alpha$) versus ([O {\scriptsize III}]/H$\beta$)  
diagnostic diagram (hereafter [O {\scriptsize I}] diagram) to classify our objects, considering the relative weakness of this line. However, we keep the results based  [O {\scriptsize I}] diagram throughout the paper for completeness. We further notice that  we consider AGN all objects meeting the requirements described above, and a large fraction of these may be LINERs. Thus, some of these objects might be powered by other processes rather than accretion (\S 1).

\begin{figure*}
\includegraphics[scale=1.3]{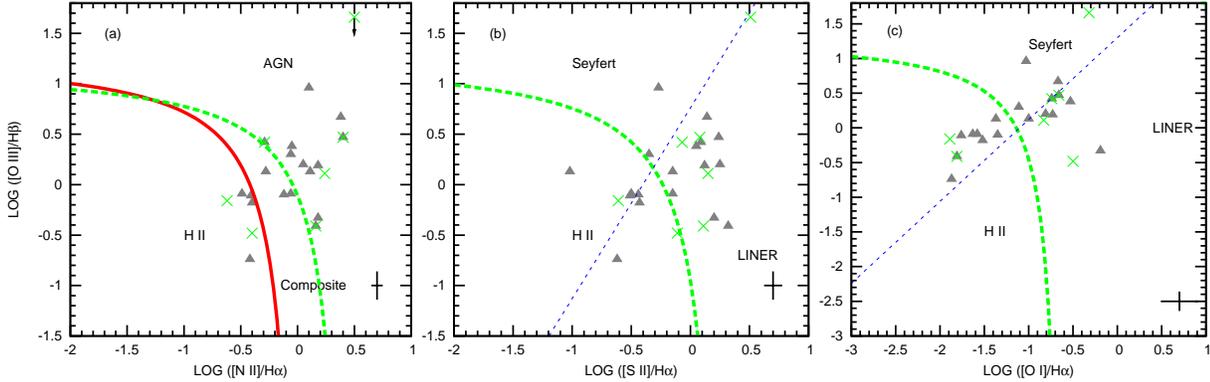}
\caption{Diagnostic diagrams of galaxies in (E+E) pairs. (a) The [N {\scriptsize II}] diagnostic diagram. (b) The  
[S {\scriptsize II}] diagnostic diagram. 
(c) The [O {\scriptsize I}] diagnostic diagram. Blue dashed line represents Seyfert/LINER line.
Elliptical galaxies can be seen on filled gray triangles, 
lenticular on green crosses and the black crosses down on the right represents the mean error data.}
\end{figure*}

\begin{figure*}
\includegraphics[scale=1.3]{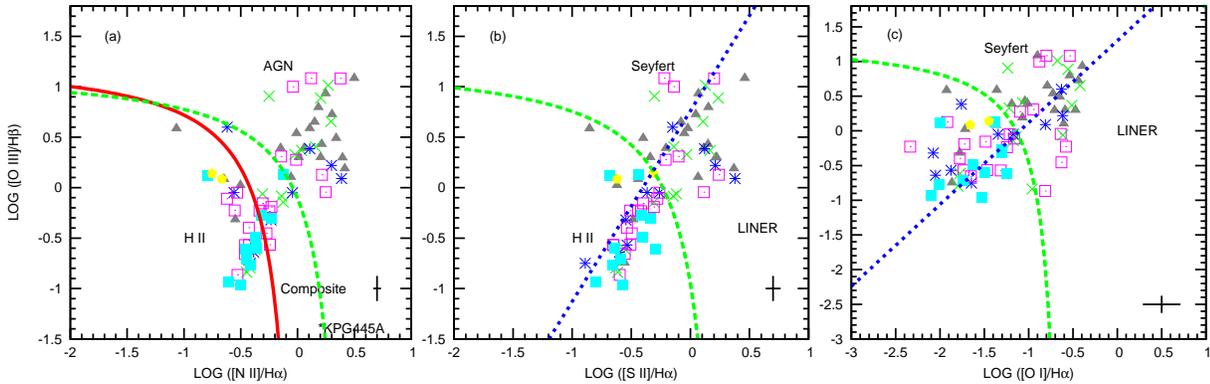}
\caption{The [N {\scriptsize II}] diagnostic diagram for (E+S) pair sample. Morphological classification is show to every object,
 elliptical galaxies can be seen on filled gray triangles, 
lenticular on green crosses, Sa on blue
 asterisks, Sb red squares with point, Sc filled cyan squares and Sm yellow pentagons. Mean error data are represented by the black cross down at right.}
\end{figure*}

In Figure 5, we present diagnostic diagrams for the (E+E) pair sample. Panel (a) shows the [N {\scriptsize II}] diagram. We
define different regions to separate galaxies with an AGN, composite galaxies (whose ionization fractions contain contributions from both an AGN  and star formation processes), and star forming galaxies (we do not distinguish  starburst from star forming galaxies, as this would require a robust measurement of the star formation rate).  We  use the ``maximum starburst line" \citep[hereafter Ke01]{Ke01},  to separate galaxies with an AGN from star forming objects (green dashed line in Fig. 5a), and the empirical line from \citet[hereafter Ka03]{Ka03}  to distinguish between pure star forming galaxies from AGN-star-forming composite objects (continuous red line).  Galaxies between the two classification lines are the  composite objects,  meaning that they contain metal-rich stellar population and AGN features in their spectra.  They consist of a circumnuclear star forming region surrounding a Seyfert or LINER nucleus. 

Panel (b) presents the[S {\scriptsize II}] diagram, that as explained above, was used to  separate between star-forming galaxies, Seyfert galaxies and LINERs. The green line corresponds to Ke01. The dashed blue line  \citep[hereafter Ke06]{Ke06} separates AGN activity between Seyfert and LINER . 
On this diagram, objects above Ke01 line can be Seyfert objects (above Ke06 line) or LINERs (objects below Ke06 line). 
Panel (c) shows  the [O {\scriptsize I}] diagram.

 Fig. 6, shows the [N {\scriptsize II}], [S {\scriptsize II}] and [O {\scriptsize I}] diagrams for the (E+S) pair sample. Similarly Figs. 7, 8, and 9 show the [N {\scriptsize II}], [S {\scriptsize II}] and [O {\scriptsize I}] 
diagnostic diagrams for the (S+S) pair sample.  
In all these plots, we use different symbols to identify galaxies with different morphological classification to study the distribution of AGN among every Hubble type. 

%Ambiguous galaxies
 %can be present in borders of lines
% separation. Galaxies that can be classified as one type of activity in one diagram can be other type in another one. So, ambiguous galaxies can be those 
%that lie in different regions for
%diagrams on Figs. 5(b)(c), 6(b)(c), 8 and 9 or composite objects that can be classified as H {\scriptsize II} or AGN activity. 

\begin{figure*}
\includegraphics[scale=1.3]{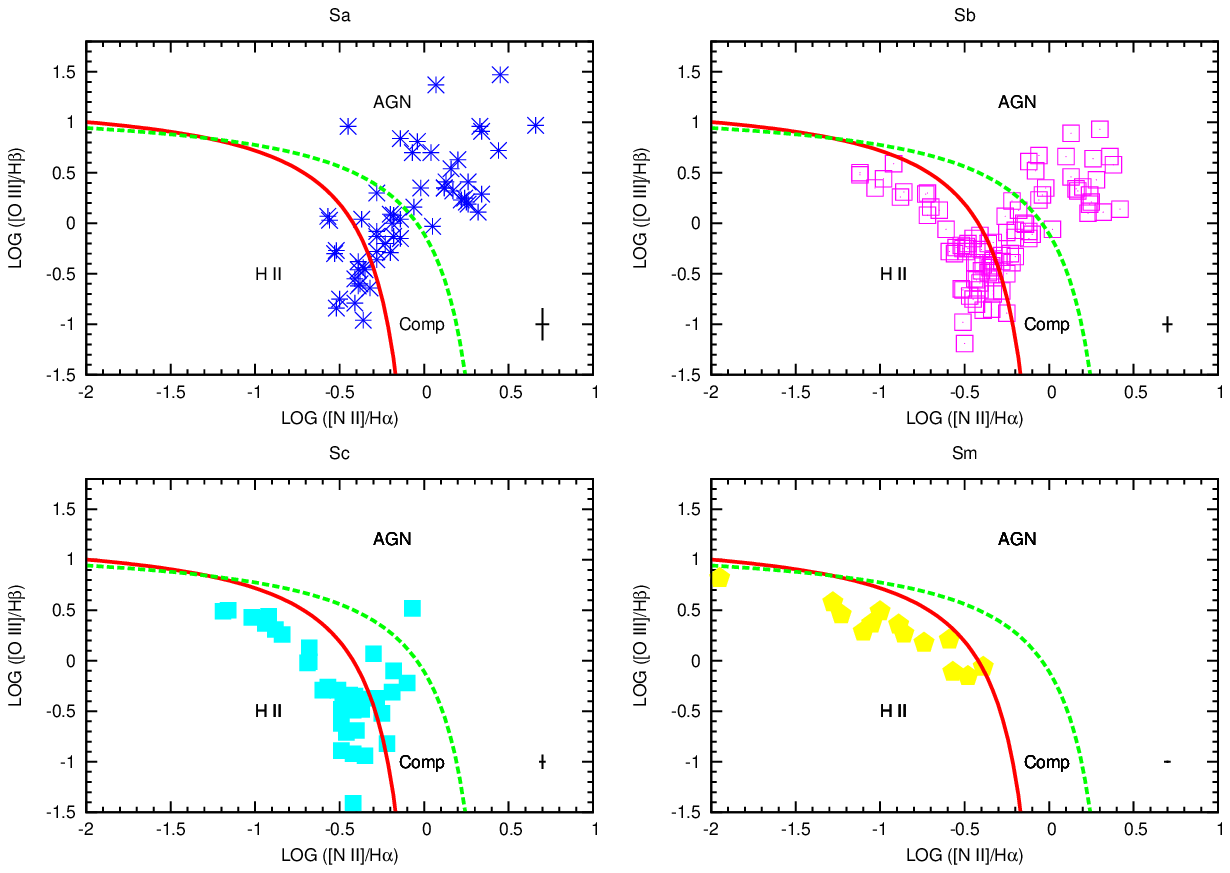}
\caption{Diagnostic diagram [N {\scriptsize II}] for galaxies in (S+S) pairs.}
\end{figure*}

\begin{figure*}
\includegraphics[scale=1.3]{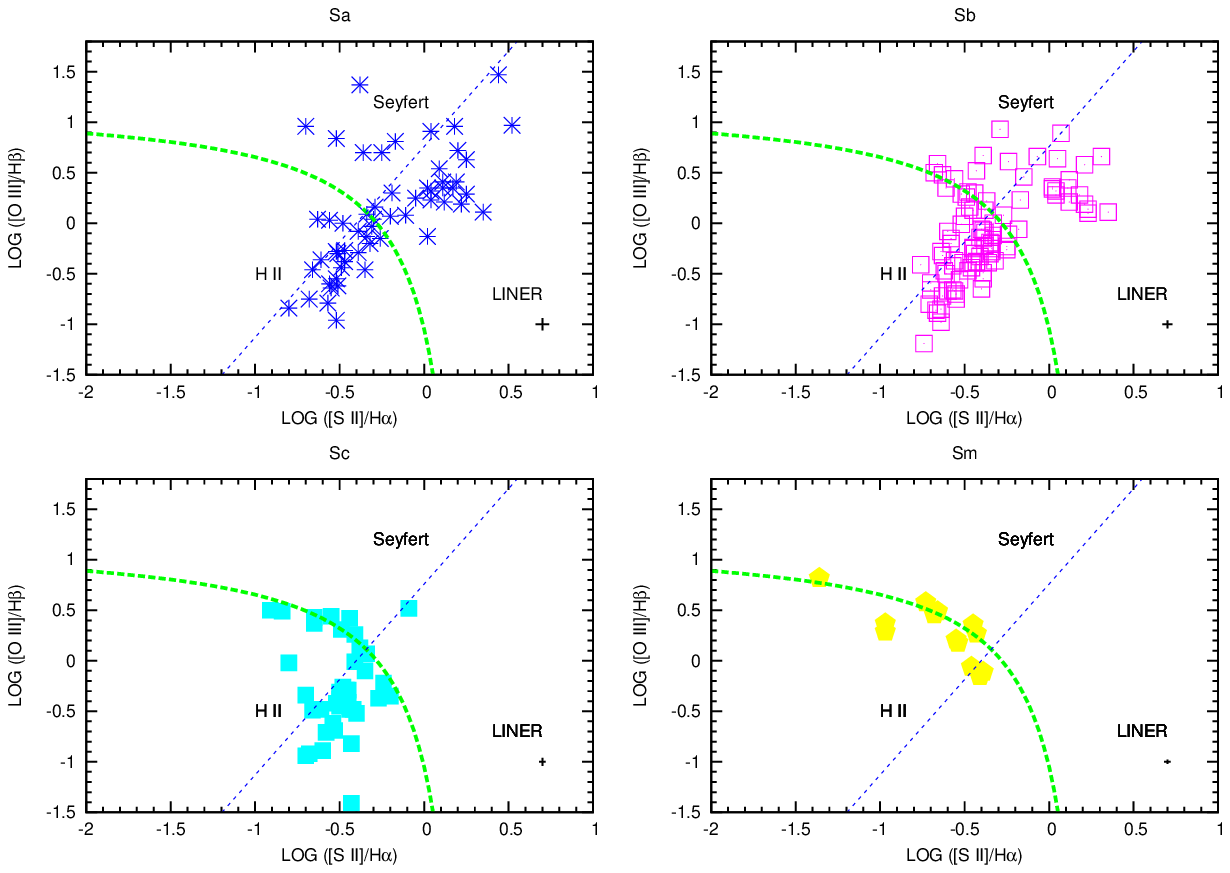}
\caption{Diagnostic diagram [S {\scriptsize II}] for galaxies in (S+S) pairs.}
\end{figure*}

\begin{figure*}
\includegraphics[scale=1.3]{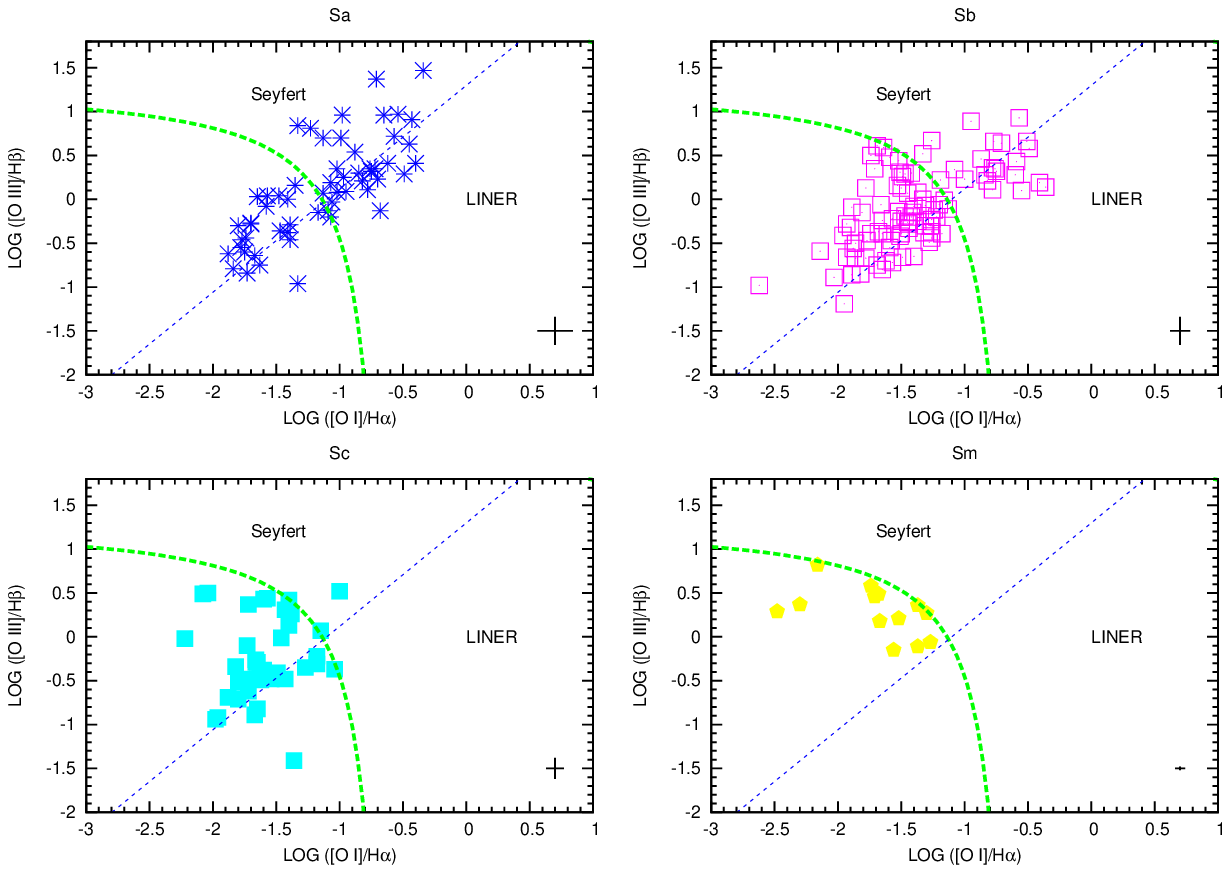}
\caption{Diagnostic diagram [O {\scriptsize I}] for galaxies in (S+S) pairs.}
\end{figure*}

 In this work, we consider composite galaxies as AGN, because they require a non-thermal continuum component to produce the level of ionization detected in their lines, that indicates the presence of  (low) accretion rates into black holes.
 This is further supported  by the fact that composite galaxies present a X-ray
 hard emission, and are considered AGN in other types of diagnostic diagrams such as the TBT diagram \citep[see][]{Trouille11}.

\subsection{Morphological Classification}

As mentioned earlier, we check the morphological classification of the galaxies in our sample. In a few cases the morphology was changed from that in the literature. These are outlined below.
We find that KPG 466B is clearly a spiral in the Sloan image and thus is considered in the (S+S) sample. According to NED, KPG 167A has elliptical 
 type, but a spiral structure is evident from the Sloan image. KPG 167B belongs to the Sb type according NED, yet it can be clearly seen as an elliptical in the SDSS.  KPG 466A presents emission lines with a double component, in both the permitted and forbidden lines. This might be due to a circumgalactic ring that can be clearly distinguished in the Sloan image.  
%We only have a morphological classification for $\sim 95$\% of the objects in our sample.
 Finally, the morphological classification of KPG 419A was obtained from a surface photometry study by \citet{Fra04} in order to avoid the existent ambiguity between NED, SIMBAD and HYPERLEDA data. We further checked the morphology of our galaxies with those classifications in the Galaxy Zoo (Lintott et al. 2008) and in the Nair \& Abraham (2010) samples. As described in  Lintott et al. (2008), we only used the morphological classifications of Galaxy Zoo with $>$80\% confidence. However, only a fraction of our galaxies were available in those catalogs (250 in Galaxy Zoo and 120 in Nair \& Abraham, with 118 being common). Comparing these classifications with the ones we use in the paper, we find an agreement for $>75\pm4\%$ of the ellipticals and S0’s, and $\sim 95\pm4\%$ of the Spirals. Interestingly, when comparing the classifications of Galaxy Zoo and Nair \& Abraham for our 118 galaxies they have in common, we find they disagree in about 10\% of them (with a larger disagreement in early types), thus stressing the difficulty of classifying the morphologies of galaxies using various methods.

\subsection{Stellar Mass of the Galaxies in the Samples }
The stellar masses were calculated using the formula described in Bell et al. (2003, ApJS 149, 289). In this case we have used 2MASS K-band fluxes (thus, a$_K=-0.283$, b$_K=0.091$), since they probe the old stellar populations which contribute the most to the stellar mass of galaxies. For the color parameter, we have used the u'-r' colors of the SDSS. The distribution of the redshift in our galaxies is 0.00052$<$z$<$0.0491 (thus they are local). Since we are using 2MASS K-band to constrain the stellar mass of the galaxies, there is no need to scale to total fluxes from SDSS.  We also note that at this band the extinction from dust is minimum. We cross-correlate our stellar masses with those presented in SDSS DR10 (the 97.5 percentile total stellar masses) which were estimated using the method described in Kauffmann et al. (2003). Since Bell et al. (2003) is using a Salpeter IMF and Kauffmann et al. (2003) is using Krupa IMF, we multiply the second by a factor of 1.5 to convert them into Salpeter.  Stellar masses calculated with both methods are very similar and they display a dispersion of 0.15dex (Bell et al. reports an expected scatter of 0.1-0.2dex). However, towards stellar masses lower than 10$^{10}$ M$_\odot$, there is a systematic overestimation of the Bell et al. (2003) method by 0.35dex. This difference could originate from the different calibrations in the estimators, but has a negligible effect in our results. 
%We were able to estimate the mass of $\sim 85$\% of the objects in our sample.

\begin{table*}\centering
    \small
    \caption{Morphology distribution and incidence of nuclear activity in (E+E) pair sub-sample derived from [N {\scriptsize II}] BPT diagnostic diagram.}
    \begin{tabular}{l l cccc l cccc}
      \hline\hline
      & \multicolumn{4}{c}{Galaxies with Emission lines}
      &&\multicolumn{4}{c}{Total sample}\\
      \cline{2-5}
      \cline{7-10}
      M.T.& Total & Star-forming & Comp & AGN+Comp && Total & Star-forming & Comp & AGN+Comp \\
      E   & 18 & 5(28\%) &  3(17\%) & 13(72\%)  && 41 &
      5(12\%) & 3(7\%)   & 13(32\%)  \\       
      S0 & 7 & 2(29\%) &  1(14\%) &  5(71\%)  && 14 &
      2(14\%) &  1(7\%)  &  5(36\%)  \\ 
          
      \hline
      Total   & 25 & 7(28\%) & 4(16\%) & 18(72\%)  && 55 &
       7(13\%)&  4(7\%) & 18(33\%)  \\
      \hline  \hline 
      \end{tabular}
      \end{table*}

 \begin{table*}\centering
    \small
    \caption{Morphology distribution and incidence of nuclear activity in (E+S) pair sub-sample derived from [N {\scriptsize II}] BPT diagnostic diagram.}
    \begin{tabular}{l l cccc l cccc}
      \hline\hline
      & \multicolumn{4}{c}{Galaxies with Emission lines}
      &&\multicolumn{4}{c}{Total sample}\\
      \cline{2-5}
      \cline{7-10}
      M.T.& Total & Star-forming & Comp & AGN+Comp && Total & Star-forming  & Comp & AGN+Comp \\
      E & 21 & 5(24\%) &  0(0\%) &  16(76\%)  && 33 &
      5(15\%) &  0(\%)  &  16(48\%)  \\ 
      S0   & 13 & 3(23\%) &  3(23\%) & 10(77\%)  && 19 &
      3(16\%) & 3(16\%)   & 10(53\%)  \\ 
      Sa & 10 & 4(40\%) & 3(30\%) & 6(60\%)  && 13 &
      4(31\%) & 3(23\%)  & 6(46\%)  \\ 
      Sb   & 21 & 8(38\%) & 6(29\%) &  13(62\%)  && 21 &
      8(38\%) & 6(29\%)  &  13(62\%)  \\ 
      Sc     & 11 & 8(73\%) & 3(27\%) &  3(27\%)  && 11 &
      8(73\%) & 3(27\%)  &  3(27\%)  \\ 
      Sm   & 2 & 2(100\%) & 0(0\%) & 0(0\%)  && 2 &
       2(100\%)&  0(0\%) & 0(0\%)  \\    
      \hline
      Total   & 78 & 30(38\%) & 15(19\%) & 48(62\%)  && 99 &
       30(30\%)&  15(15\%) & 48(48\%)  \\
      \hline  \hline 
      \end{tabular}
      
      \end{table*}

\begin{table*}\centering
    \small
    \caption{Morphology distribution and incidence of nuclear activity in (S+S) pair sub-sample derived from [N {\scriptsize II}] BPT diagnostic diagram.}
    \begin{tabular}{l l cccc l cccc}
      \hline\hline
      & \multicolumn{4}{c}{Galaxies with Emission lines}
      &&\multicolumn{4}{c}{Total sample}\\
      \cline{2-5}
      \cline{7-10}
      M.T.& Total & Star-forming & Comp & AGN+Comp && Total & Star-forming  & Comp & AGN+Comp \\
      Sa & 60 & 22(37\%) & 12(20\%) & 36(60\%)  && 65 &
      22(34\%) & 12(18\%)  & 36(55\%)  \\ 
      Sb   & 92 & 54(59\%) & 16(17\%) &  38(41\%)  && 97 &
      54(56\%) & 16(16\%)  &  38(39\%)  \\ 
      Sc     & 39 & 33(85\%) & 6(15\%) &  6(15\%)  && 40 &
      33(83\%) & 6(15\%)  &  6(15\%)  \\ 
      Sm   & 12 & 11(92\%) & 1(8\%) & 1(8\%)  && 12 &
       11(92\%)&  1(8\%) & 1(8\%)  \\    
      \hline
      Total   & 203 & 120(59\%) & 35(17\%) & 83(40\%)  && 214 &
       120(56\%)&  35(16\%) & 83(39\%)  \\
      \hline  \hline 
      \end{tabular}
      \end{table*}

\section{Results}

\begin{table*}\centering
      \small
      \caption{Incidence type of nuclear activity for different morphologies from [S {\scriptsize II}] and [O {\scriptsize I}] diagnostic diagrams for (E+E) sample.}    
      \begin{tabular}{l l cccc l cccc}
      \hline\hline
      & \multicolumn{4}{c}{ [S {\scriptsize II}] Diagram}
      &&\multicolumn{4}{c}{[O {\scriptsize I}] Diagram}\\
      \cline{2-5}
      \cline{7-10}
    M.T.& Total & AGN & Seyfert & LINER && Total & AGN & Seyfert & LINER \\
    E & 18 & 12(67\%) & 2(11\%) & 10(56\%) && 18 & 10(56\%) & 5(50\%) & 5(50\%)\\
    S0 & 7 & 5(71\%) & 0(0\%) & 5(100\%) && 7 & 5(71\%) & 1(20\%) & 4(80\%)\\
    \hline
      Total & 25 & 17(68\%)& 2(12\%)& 15(88\%)  && 25 & 15(60\%) & 6(40\%) & 9(60\%)\\
      \hline  \hline 
      \end{tabular}
      \end{table*}
\begin{table*}\centering
      \small
      \caption{Incidence type of nuclear activity for different morphologies from [S {\scriptsize II}] and [O {\scriptsize I}] diagnostic diagrams for (E+S) sample.}    
      \begin{tabular}{l l cccc l cccc}
      \hline\hline
      & \multicolumn{4}{c}{ [S {\scriptsize II}] Diagram}
      &&\multicolumn{4}{c}{[O {\scriptsize I}] Diagram}\\
      \cline{2-5}
      \cline{7-10}
      M.T.& Total & AGN & Seyfert & LINER && Total & AGN & Seyfert & LINER \\
      E & 21 & 16(76\%) &  5(31\%) &  11(69\%)  && 21 &
      16(76\%) &  10(63\%)  &  6(37\%)  \\ 
      S0   & 13 & 9(69\%) &  2(22\%) & 7(78\%)  && 13 &
      8(62\%) & 5(63\%)   & 3(37\%)  \\ 
      Sa & 10 & 4(40\%) & 1(25\%) & 3(75\%)  && 10 & 3(30\%) & 1(33\%) & 2(67\%) \\ 
      Sb   & 21 & 7(33\%) & 2(29\%) &  5(71\%)  && 21 &
      9(43\%) & 5(56\%)  &  4(44\%)  \\ 
      Sc     & 11 & 0(0\%) & 0(0\%) &  0(0\%)  && 11 & 0(0\%) & 0(0\%) &  0(0\%)\\ 
      Sm   & 2 & 0(0\%) & 0(0\%) & 0(0\%)  && 2 & (0\%) & 0(0\%) & 0(0\%)  \\    
      \hline
      Total   & 78 & 36(46\%) & 10(28\%) & 26(72\%)  && 78 &
       36(46\%)&  21(58\%) & 15(42\%)  \\
      \hline  \hline 
      \end{tabular}
      \end{table*}
\begin{table*}\centering
      \small
      \caption{Incidence type of nuclear activity for different morphologies from [S {\scriptsize II}] and [O {\scriptsize I}] diagnostic diagrams for (S+S) sample.}    
      \begin{tabular}{l l cccc l cccc}
      \hline\hline
      & \multicolumn{4}{c}{ [S {\scriptsize II}] Diagram}
      &&\multicolumn{4}{c}{[O {\scriptsize I}] Diagram}\\
      \cline{2-5}
      \cline{7-10}
      M.T.& Total & AGN & Seyfert & LINER && Total & AGN & Seyfert & LINER \\
      Sa & 60 & 29(48\%) & 7(24\%) & 22(76\%)  && 60 & 31(52\%) & 18(58\%) & 13(42\%) \\ 
      Sb   & 92 & 30(33\%) & 13(43\%) &  17(57\%)  && 92 &
      23(25\%) & 10(43\%)  &  13(57\%)  \\ 
      Sc     & 39 & 5(13\%) & 4(80\%) &  1(20\%)  && 39 & 2(5\%) & 2(100\%) &  0(0\%)\\ 
      Sm   & 12 & 1(8\%) & 1(100\%) & 0(0\%)  && 12 & (0\%) & 0(0\%) & 0(0\%)  \\    
      \hline
      Total   & 203 & 69(34\%) & 29(42\%) & 40(58\%)  && 203 &
       56(28\%)&  30(54\%) & 26(46\%)  \\
      \hline  \hline 
      \end{tabular}
      \end{table*}

Tables 1 to 3, show our results using the [N {\scriptsize II}] diagnostic diagram for the (E+E), (E+S), and (S+S) pairs of galaxies, respectively. These Tables present the fraction of AGN, composite, and star formation objects as a function of morphological type.  The type of nuclear activity (LINER or Seyfert), obtained with the  [S {\scriptsize II}] and  [O {\scriptsize I}] diagnostic diagrams for the different samples of paired galaxies, are shown in Tables 4, 5, and 6.
Finally, Table 7 summarizes our main findings in an easy and comparative way, showing the fraction of AGN as a function of morphological type of each object for our three samples of pairs. In this Table we also present the results from the sample of isolated galaxies by  H-I13, for comparison. 

Our results indicate that 48\% of the emission line galaxies in pairs show the presence of nuclear (non-thermal) activity. When galaxies that do not show emission lines (according to the quantitative definition given in \S 2)  are included in the calculations, this frequency decreases to 40\%.
We can compare this incidence with the sample of isolated galaxies by  H-I13. Isolated objects with emission lines present nuclear activity in $\sim$    43\% of the cases, and $\sim$41\% for the total sample (again, including non-emission galaxies; we note here that the definition of non-emission galaxies follows a self-consistent approach in this paper and in H-I13). This difference is clearly not statistically significant, indicating that interactions are not a sufficient condition in producing nuclear activity. However, it is important to note that these results are based on our complete sample, and kind of interaction in the pairs (S+S, E+E, or E+S), as well as morphological and mass distributions among the different samples are different. Thus, the results on the total sample might be leaded by these different properties. One of the aims of the present work is precisely to separate these different effects. 

Table 7 presents our results as a function of type of interaction for the different morphological classification of the objects. It can be observed that late type galaxies in (S+S) pairs have similar frequency of AGN than isolated galaxies. However, spirals in (E+S) pairs type galaxies have a frequency $\sim$10 \% larger than (S+S) and isolated galaxies. These results do not change if only  galaxies with emission lines or  the total samples are considered. The difference between the (E+S) pairs and the isolated and (S+S) pairs is of the order of the Poisson error in the measurements (10\%), so not significative. As can be observed in the Table, the difference is mainly driven by the larger fraction of AGN in Sb galaxies that are members of the (E+S) pairs, that host an AGN 15-20\% more often than Sb galaxies in the other groups. However, due to the smaller number statistics, the Poisson error on this sample is similar to the difference.
Observing at early type galaxies with emission lines in (E+S) and (E+E) pairs, we can observe that they have a high fraction of AGN in their nuclei, but comparable with the frequency observed in isolated galaxies. When galaxies without emission features are considered, the fraction decreases, but remains similar among the samples, with the only difference that galaxies in (E+E) pairs seem to have a $\sim~2\sigma$ (17-18\%) smaller fraction of active galaxies.

Overall, it is clear from Table 7 that the main factor in the presence of AGN in galaxies is the morphological type. If only emission line objects are considered, Early type galaxies have the largest fraction, decreasing monotonically as the galaxies are of later type.  However, when all objects are taken into account, the fraction of AGN in early types decreases, and the largest fraction of AGN are found in Sa and Sb galaxies. 

As shown by H-I13 the mass of the galaxies is another dominant factor in determining the AGN fraction in galaxies. We find the same trend in our samples of paired galaxies. As can be seen in Figure 4, for galaxies with masses larger than 10$^{10.5}$M$_\odot$, the fraction of galaxies hosting an active nucleus raises above 50\%. For masses larger than  10$^{11}$M$_\odot$, nearly all objects have an AGN component. Given that there is some relation between the morphological and the mass distributions in galaxy samples this result is not entirely surprising. However, there is more to it than this simple link between morphology and mass.  A direct comparison  (independent on morphology) suggests that paired galaxies in massive objects tend to host AGN more often than isolated galaxies at a single mass value, as observed in Figure 10. However, when one takes into account both the morphology and mass distribution in the samples this strong difference vanishes. This is observed in Figure 11, where the fraction of AGN is plotted for early and late type galaxies (as function of type of interaction) for the total sample and for objects with masses 10$^{10.0}$M$_\odot$ and 10$^{10.5}$M$_\odot$. For this plot, only objects with emission lines are considered, and to avoid small number statistics, all spirals have been grouped as Late type objects, while E and S0 have been grouped as early type. It is clear from this Figure, that any differences in AGN fraction for late type objects are of the 10\% (similar to the error), and that early type objects show strikingly the same fraction independently of the environment and type of interaction. In fact, from this plot it is clear that early type galaxies showing emission lines are almost all AGN, and this fraction decreases towards later types. We note however, that when galaxies without emission lines are considered, early type galaxies show a lower fraction of AGN when compared to later types (as found for the entire sample), even at high masses.

%Sc and Sm galaxies show practically only H {\scriptsize II} like regions. As we can see in Tables 5, 6 and 7 for every galaxy, the morphological 
%classification according to each  BPT diagram taking in account the errors in data. 
%COMP-H {\scriptsize II} means that galaxy is classified like AGN but can be H {\scriptsize II} by error in measurement, and the same thing for LINER-Seyfert (L-S),
 %Seyfert-LINER (S-L) or other combination. (See Tables 8,9 and 10 at the end of this article where we show logarithm intensities ratios and 
%type classification for every object in the three subsamples).

 The most striking result in our analysis is that in all the 385 revised spectra, only 4 objects present AGN type 1 activity, which represents $\sim$ 1$\%$ of the sample. This result does not seems to support the unified model at least in its simpler version.

\begin{figure}
\includegraphics[scale=0.65]{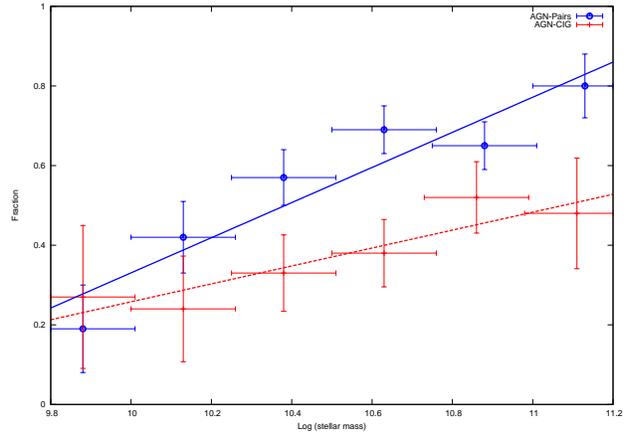}
\caption{Comparison between fraction of AGN in pairs and AGN isolated galaxies as function
of stellar mass. AGN in pairs are represented by (blue) filled
circles and AGN isolated objects by (red) crosses. Errors
in the y-direction are the standard deviation per bin and
“errors” in x-direction simply denote the full range of
mass in each bin. A different slope is remarkable above 10$^{10}$ solar masses between AGN in pairs and AGN in isolated systems. However, such slope might be the result of the different morphological mixes in both samples.}
\end{figure}

\section{Discussion}	
  
%\begin{onecolumn}
\begin{figure*}
%\begin{subfigure}{.5\textwidth}
\centering
\includegraphics[scale=0.28]{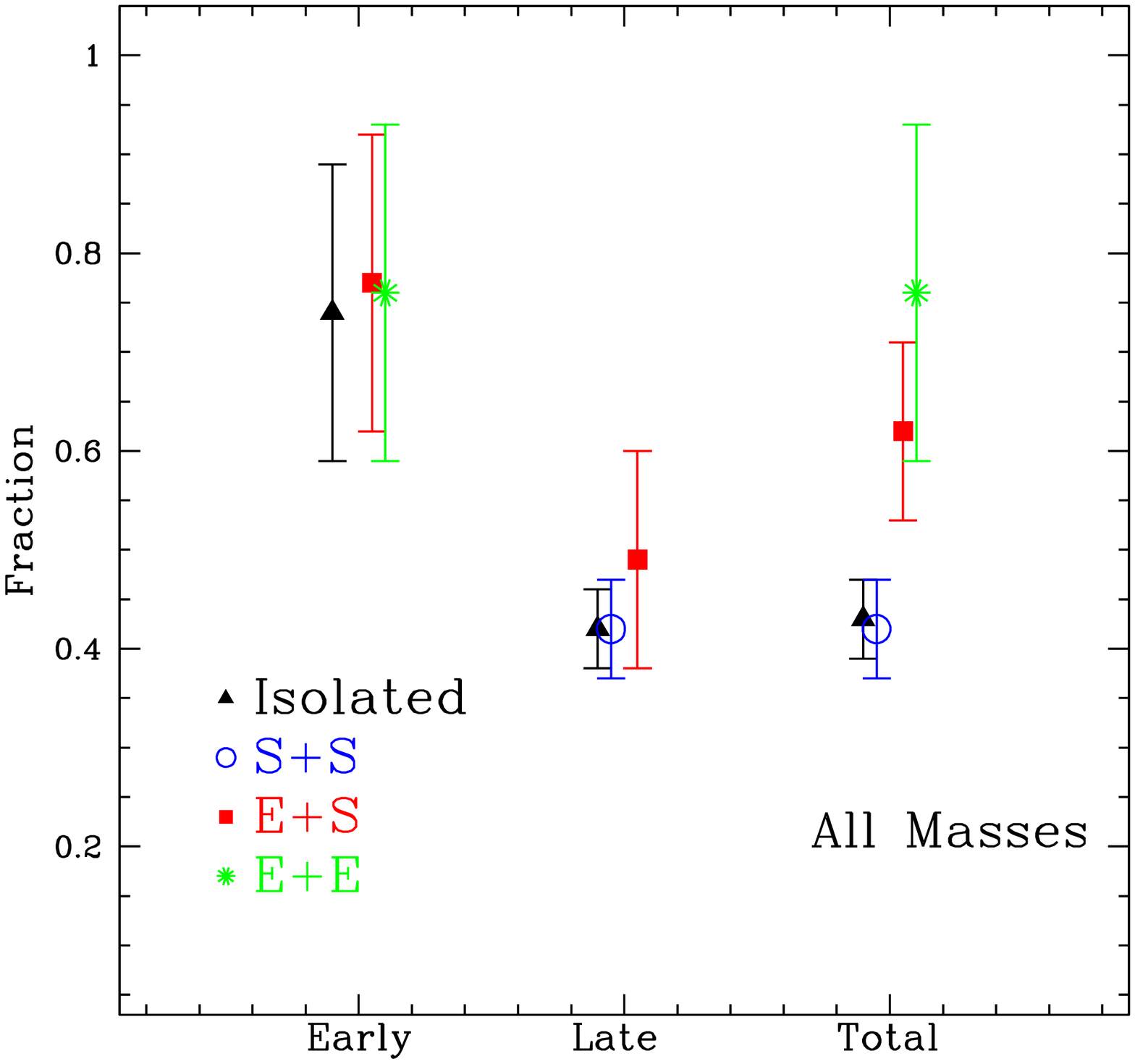}
\centering
\includegraphics[scale=0.28]{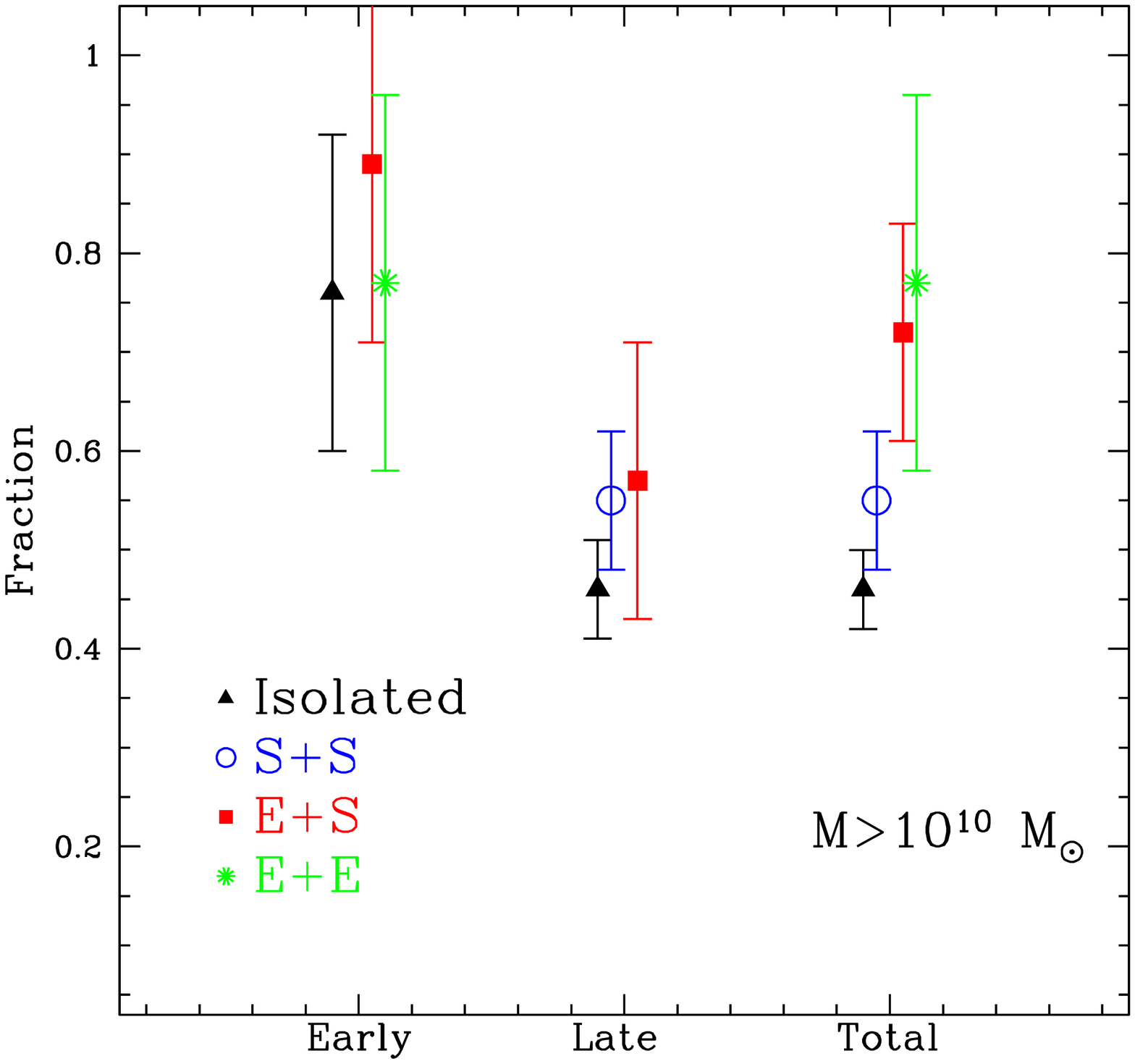}
%\end{subfigure}%
%\begin{subfigure}{.5\textwidth}
\centering
\includegraphics[scale=0.28]{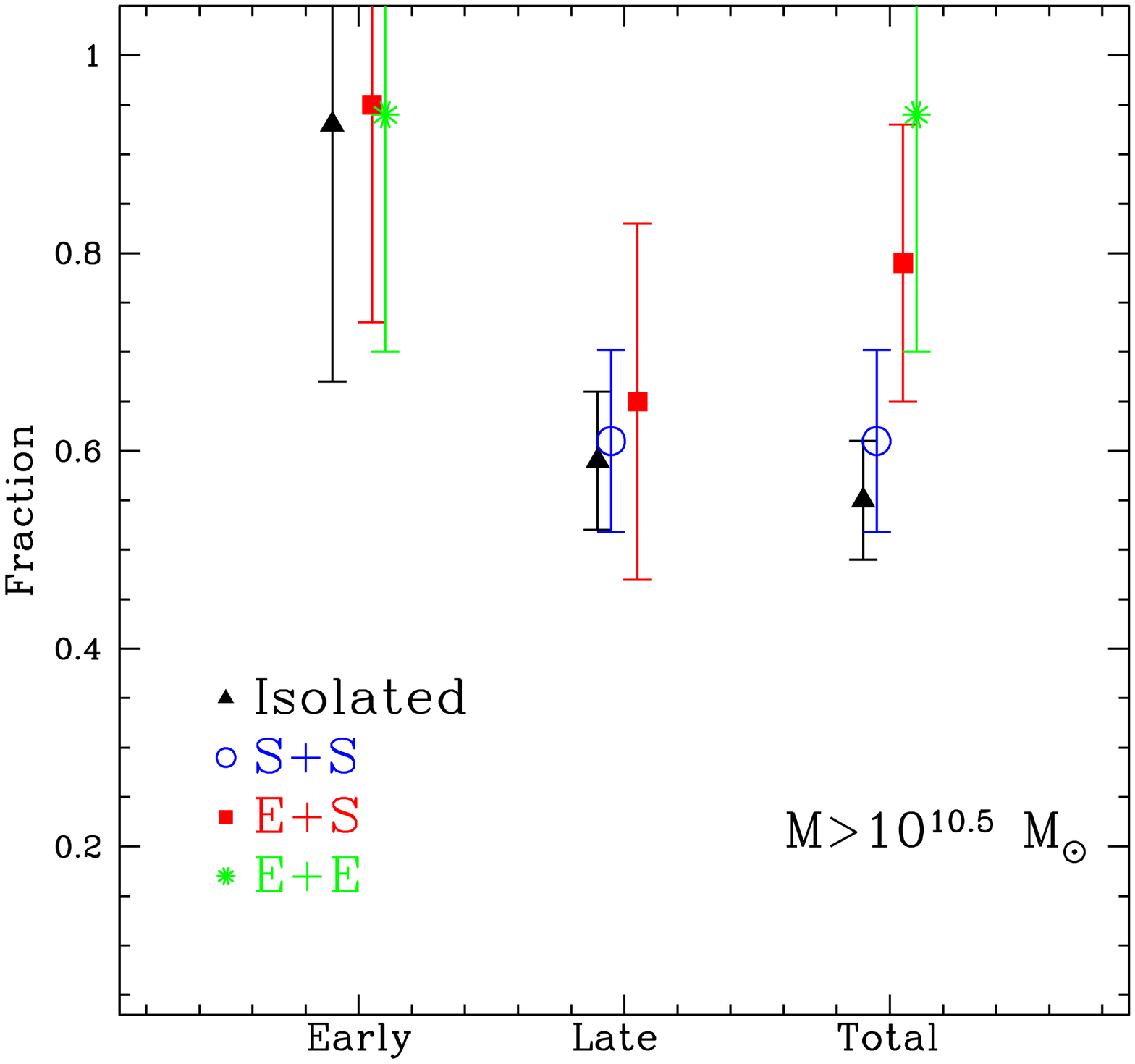}
%\end{subfigure}%
\caption{Fraction of AGN, considering only galaxies with emission lines, in the different type of pairs and in isolated galaxies as a function of morphological type for three different ranges of mass: (Left) All galaxies; (Center) Galaxies with mass $>10^{10}$M$_\odot$; and (Right) Galaxies with mass $>10^{10.5}$M$_\odot$. No significant excess of AGN in interacting galaxies is observed when galaxies with similar masses and morphologies are considered.}
\end{figure*}

We have made a comparison of the incidence of nuclear activity in a sample of paired galaxies with the incidence in a sample of isolated galaxies. 
All our samples were analyzed in a homogeneous and consistent way. This comparison has been done separating according to morphological types and stellar masses.

Our results confirm that there is a link between morphology and nuclear activity. When galaxies with and without emission lines are considered (the entire samples), early type spirals host the larger fractions of AGN. The incidence of activity shows a growing distribution from E to Sb morphological types, decreasing sharply for later types. When only galaxies with emission lines are considered, the incidence of AGN activity in Elliptical
 and Spheroidal galaxies is particularly large (around $\sim$75 \%, almost all LINERs). This incidence is similar in both isolated and paired galaxies. This means that if an early type galaxy shows emission lines, it is very likely that it hosts nuclear activity. For spiral galaxies with emission lines (both paired and isolated), the incidence of AGN activity (versus star forming objects)
decreases gradually form early to late type galaxies, in other words, as the size of the bulge decreases. 
Similar results have been found 
 by \citet {2004A&A...420..873V, 2011RMxAA..47..361C, Sa12, 2012arXiv1206.6777H}. 

The fraction of AGN in spirals can be readily understood considering that both the presence of a dominant bulge and a large reservoir of gas are required to canalize gas and feed a supermasive black hole (SMBH). The large fraction of AGN in Ellipticals when only emission line objects are taken into account may be explained from the fact that, when these galaxies have some supply of gas (either its own, or accreted from the companion), it can reach the inner regions easily and feed the SMBH (rather than stop in the way to the center and produce star formation). When non-emission line galaxies are also considered, the fraction of AGN in early types decreases significantly, since there is a larger number of non-emitting sources in this samples, considering that in early galaxies there is negligible ongoing star formation.  The fraction of Seyfert nuclei increases from early to late types. Only $\sim 4$\% of the galaxies in (E+E) pairs host a Seyfert nuclei ($\sim12$\% of the AGN in (E+E) pairs). However, the fraction of Seyfert nuclei rises to $\sim 10$\% in (E+S) pairs  ($\sim28$\% of the AGN in this kind of pairs), and to $\sim 14$\% in (S+S) systems  ($\sim42$\% of the AGN in (s+S) pairs). 
These trends may result from the limited supply of material in early type objects.

A striking and surprising result is that isolated and paired galaxies yield similar levels of nuclear activity incidence, with only marginal indication for an enhancement in interacting, late type objects. The lack of differences appears to be independent of mass. This suggests that this type of relatively weak (i.e. tidal) interaction between low luminosity galaxies is not a dominant  triggering force of nuclear activity, at least, for low luminosity AGN (LLAGN), as the ones found in the pairs of galaxies and in isolated objects. Although there is clear evidence that more luminous AGN can be triggered by interactions (see \S 1), the latter are clearly not a sufficient condition for the onset of a SMBH.  Perhaps, stronger gravitational interactions (such as a very close encounter and of course mergers) are needed for gravitational triggering of activity. 

Interactions are not a necessary condition either for the existence of these LLAGN, since the fraction of isolated galaxies hosting  such a nucleus is large and comparable to pairs of galaxies. Taken at face value, our results indicate that, although non-axysimetric perturbations (such as bars or interactions) are important in AGN triggering,  the exact necessary and sufficient processes to form an active galaxy (either of high or low luminosity) remain a mystery.  Overall, these results indicate that secular evolution processes (see H-I13) can trigger/maintain low luminosity AGN activity, and as such might be important in some processes of the galaxy. However, as discussed by H-I13, AGN with such low luminosities and accretion rates ($0.001~- ~10^{-5}$ M$_\odot$/yr ) galaxies cannot grow their central black holes through these low-level activity, nor can they evolve from one type of object to another.

Perhaps, the more  relevant result from this work is the fact that out of the 150 AGN galaxies in pairs, only 4 have (albeit a small) broad component:
We find one 1.9 Sy, two 1.8 Sy and one 1.5 Sy. Not even one type 1 Sy.  This result is at odds with the simplest formulation of the unified model (UM) \citep{1993ARA&A..31..473A}, which takes into account orientation and obscuration effects only. An evolutionary scheme, on the other hand, may explain this result. A Seyfert 1 may require as much as 1 Gyr after an interaction to appear  (Krongold  et  al.  2002). In this case, an  interaction  may trigger  first  a  circumnuclear  starburst,  and subsequently  non-thermal  nuclear  activity. 
Such scheme could explain the lack of type 1 nucleus in these pairs of galaxies, considering that there has not been
enough time for the Sy 1 nucleus to appear. However, the lack of enhancement in the fraction of AGN in the pairs of galaxies, when compared to isolated objects, would require that we are seeing these pairs on its initial approach. Whether this is the case or not, our results are consistent with the idea that the BLR in active galaxies can only be formed at higher luminosities/accretion rates (e.g. Nicastro 2000, Elitzur \& Ho 2009).

\section*{Acknowledgments}
FJHI acknowledges a graduate student scholarship from
CONACYT. DD acknowledge support from grant IN111610 PAPIIT, UNAM.

%\end{onecolumn}

\begin{table*}\centering
    \scriptsize
    \caption{Incidence of nuclear activity for the different morphological types as a function of pair morphology derived from [N {\scriptsize II}] BPT diagnostic diagram.}
    \begin{tabular}{l c cccc c cccc}
      \hline\hline
      & \multicolumn{8}{c}{Fraction of AGN (including composite objects) in Early Type Galaxies}\\
      & \multicolumn{4}{c}{Galaxies with Emission lines}
      &&\multicolumn{4}{c}{Total sample}\\
      \cline{2-5}
      \cline{7-10}
      M.T. &  E+E  & E+S & S+S &  Isolated & & E+E  & E+S & S+S &  Isolated  \\
            \cline{2-5}
      \cline{7-10}
      E   & 13/18 (72\%) & 16/21 (76\%) &  -- & 10/14 (71\%) &     & 13/41 (32\%) & 16/33 (48\%) &   --   & 10/23 (43\%)  \\       
      S0 &   5/7 (71\%)   & 10/13 (77\%) &  -- & 21/30 (70\%) &     &   5/14 (36\%) & 10/19 (53\%) &   --   & 21/38 (55\%)  \\ 
          
      \hline
      Total  & 18/25 (72\%)&  26/34 (76\%) & -- & 31/44 (70\%)  & &  18/55 (33\%)&  26/52 (50\%) &  --   & 31/61 (51\%)    \\
      \hline  \hline 
    & \multicolumn{8}{c}{Fraction of AGN (including composite objects) in Late Type Galaxies}\\
      & \multicolumn{4}{c}{Galaxies with Emission lines}
      &&\multicolumn{4}{c}{Total sample}\\
     \cline{2-5}
      \cline{7-10}
      M.T. &  E+E  & E+S & S+S &  Isolated & & E+E  & E+S & S+S &  Isolated  \\
            \cline{2-5}
      \cline{7-10}
     Sa     &  -- &  6/10 (60\%)  &  36/60 (60\%) &  25/38 (66\%) &     &   -- & 6/13 (46\%) &   36/65 (55\%)   & 25/38 (66\%)  \\ 
     Sb     &  -- & 13/21 (62\%) &  38/92 (41\%) &55/122 (45\%) &     &   -- &13/21 (62\%) &  38/97 (39\%)   &55/124 (44\%)  \\ 
     Sc     &  -- &   3/11 (27\%) &    6/39 (15\%) &37/124 (30\%) &     &   -- &  3/11 (27\%) &    6/40 (39\%)   &37/124 (30\%)  \\ 
     Sm     &  -- &   0/2  (0\%)   &    1/12 (8\%)   &  2/20  (10\%) &     &   -- &   0/2  (0\%)   &    1/12  (8\%)    & 2/20 (10\%)  \\ 
      \hline
      Total  &  -- &  22/44 (50\%) &81/203 (40\%)& 119/304 (39\%)  & &  -- &  22/47 (47\%) &81/214 (38\%)& 119/306 (39\%)     \\
      \hline  \hline 
    & \multicolumn{8}{c}{Fraction of AGN (including composite objects) in the Pair Samples}\\
      & \multicolumn{4}{c}{Galaxies with Emission lines}
      &&\multicolumn{4}{c}{Total sample}\\
     \cline{2-5}
      \cline{7-10}
       &  E+E  & E+S & S+S &  Isolated & & E+E  & E+S & S+S &  Isolated  \\
     \cline{2-5}
      \cline{7-10}
      Total  &  18/25 (72\%)  &  48/78 (62\%) &81/203 (40\%)& 150/348 (43\%)  & &  18/55 (33\%) &  48/99 (48\%) &81/214 (38\%)& 150/367 (41\%)     \\
      \hline  \hline 
    & \multicolumn{8}{c}{Total Fraction of AGN in Pairs vs. Isolated Samples}\\
      & \multicolumn{4}{c}{Galaxies with Emission lines}
      &&\multicolumn{4}{c}{Total sample}\\
     \cline{2-5}
\cline{7-10}
& \multicolumn{2}{c}{Galaxies in Pairs} & \multicolumn{2}{c}{Isolated Galaxies}
      && \multicolumn{2}{c}{Galaxies in Pairs} & \multicolumn{2}{c}{Isolated Galaxies}\\
      \cline{2-5}
      \cline{7-10}
      Total  &  \multicolumn{2}{c}{147/306(48\%)} & \multicolumn{2}{c}{150/348 (43\%)}  & &  \multicolumn{2}{c}{147/368(40\%)} & \multicolumn{2}{c}{150/367 (41\%)}   \\
      \hline  \hline

      \end{tabular}
      \end{table*}

\clearpage

\appendix
\section{Samples of (E+E), (E+S), and (S+S) Pairs}

 Tables \ref{a1}1,  \ref{a2}2, and \ref{a3}3 present the samples for the different types of pairs.

\begin{deluxetable}{ccccccc} \tabletypesize{\footnotesize}
\label{a1}
\tablecolumns{7} \tablewidth{0pt} \tablecaption{Logarithm intensities
ratios and their errors for (E+E) sample.} 
\tablehead{ \colhead{Object}			 &
\colhead{M.T.\,\tablenotemark{a}}  	 &  \colhead{LOG([N
II]/H$\alpha$)}      &  \colhead{LOG([O {\scriptsize III}]/H$\beta$)} 	 &
\colhead{LOG([S {\scriptsize II}]/H$\alpha$)}      & \colhead{LOG([O {\scriptsize I}]/H$\alpha$) }
& \colhead{Type }	 } \startdata \sidehead{Elliptical} \hline

KPG145B	&	E	&	0.1852	$\pm$	0.0233	&	0.1772	$\pm$	0.0714	&	0.2792	$\pm$	0.0292	&	-0.7272	$\pm$	0.0758	&	LINER	\\ 
KPG170B	&	E	&	0.1994	$\pm$	0.0698	&	0.0503	$\pm$	0.1249	&	0.2482	$\pm$	0.0792	&	-0.8132	$\pm$	0.2803	&	LINER	\\ 
KPG192B	&	E	&	0.3005	$\pm$	0.0333	&	0.2032	$\pm$	0.1058	&	0.2222	$\pm$	0.0445	&	-0.4733	$\pm$	0.0740	&	LINER	\\ 
KPG204A	&	E	&	-0.6887	$\pm$	0.0087	&	-0.2983	$\pm$	0.0685	&	-0.5590	$\pm$	0.0159	&	-1.6857	$\pm$	0.0682	&	Star-forming	\\ 
KPG223A	&	E	&	0.3833	$\pm$	0.0353	&	-0.0495	$\pm$	0.0947	&	0.1936	$\pm$	0.0389	&	-0.5300	$\pm$	0.0684	&	LINER	\\ 
KPG235B	&	E	&	0.2970	$\pm$	0.0240	&	-0.0619	$\pm$	0.0803	&	-0.1959	$\pm$	0.0396	&	-1.1057	$\pm$	0.1250	&	Sy2	\\ 
KPG238A	&	E	&	-0.0985	$\pm$	0.0713	&	-0.1181	$\pm$	0.1809	&	-0.3106	$\pm$	0.1496	&	-1.3532	$\pm$	0.6396	&	AGN	\\ 
KPG238B	&	E	&	-0.2778	$\pm$	0.0091	&	0.1318	$\pm$	0.0065	&	-1.0212	$\pm$	0.0210	&	-1.3708	$\pm$	0.0233	&	AGN*	\\ 
KPG361B	&	E	&	0.1256	$\pm$	0.0889	&	0.1128	$\pm$	0.2682	&	0.0330	$\pm$	0.1426	&	-1.0029	$\pm$	0.5114	&	S-L	\\ 
KPG372B	&	E	&	-0.0923	$\pm$	0.0071	&	-0.4935	$\pm$	0.0085	&	-0.3473	$\pm$	0.0071	&	-1.6329	$\pm$	0.0191	&	Star-forming	\\ 
KPG373A	&	E	&	0.5639	$\pm$	0.0450	&	0.2682	$\pm$	0.1039	&	0.1240	$\pm$	0.0695	&	-0.6585	$\pm$	0.1421	&	S-L	\\ 
KPG399A	&	E	&	0.6728	$\pm$	0.0459	&	0.3764	$\pm$	0.2281	&	0.2738	$\pm$	0.0657	&	-0.6724	$\pm$	0.1656	&	S-L	\\ 
KPG454A	&	E	&	-0.7423	$\pm$	0.0065	&	-0.4240	$\pm$	0.0199	&	-0.4600	$\pm$	0.0078	&	-1.8731	$\pm$	0.0401	&	Star-forming	\\ 
KPG454B	&	E	&	-0.1145	$\pm$	0.0110	&	-0.4070	$\pm$	0.0196	&	-0.3490	$\pm$	0.0133	&	-1.7599	$\pm$	0.0843	&	Star-forming	\\ 
KPG489A	&	E	&	-0.3270	$\pm$	0.2184	&	0.1827	$\pm$	0.2343	&	0.3003	$\pm$	0.2657	&	-0.1871	$\pm$	0.3307	&	LINER	\\ 
KPG489B	&	E	&	-0.0934	$\pm$	0.0477	&	-0.0634	$\pm$	0.1344	&	-0.0082	$\pm$	0.0654	&	-1.5781	$\pm$	0.7487	&	AGN	\\ 
KPG510B	&	E	&	 0.1006	$\pm$	0.0070	&	0.9620	$\pm$	0.0095	&	-0.2733	$\pm$	0.0095	&	-1.0261	$\pm$	0.0205	&	Sy1.8	\\ 
KPG600B	&	E	&	-0.1817	$\pm$	0.0169	&	-0.4024	$\pm$	0.0415	&	-0.2919	$\pm$	0.0217	&	-1.5216	$\pm$	0.1035	&	Star-forming	\\ 

\sidehead{Spheroidal} \hline 
KPG184B	&	S0	&	0.4740	$\pm$	0.0313	&	0.3984	$\pm$	0.1213	&	0.2363	$\pm$	0.0469	&	-0.6619	$\pm$	0.1102	&	LINER	\\ 
KPG208A	&	S0	&	0.4751	$\pm$	0.0102	&	-0.3961	$\pm$	0.0124	&	-0.1139	$\pm$	0.0107	&	-0.5039	$\pm$	0.0093	&	Sy1.5	\\
KPG208B	&	S0	&	-0.1639	$\pm$	0.0068	&	-0.6217	$\pm$	0.0095	&	-0.4434	$\pm$	0.0073	&	-1.8897	$\pm$	0.0319	&	Star-forming	\\ 
KPG277B	&	SB0	&	0.1127	$\pm$	0.0388	&	0.2444	$\pm$	0.1296	&	0.1511	$\pm$	0.0584	&	-0.8281	$\pm$	0.1667	&	LINER	\\ 
KPG359A	&	SB0	&	0.4238	$\pm$	0.0893	&	-0.2928	$\pm$	0.2166	&	0.0869	$\pm$	0.0805	&	-0.7440	$\pm$	0.1854	&	S-L	\\ 
KPG431B	&	S0	&	-0.4124	$\pm$	0.0725	&	0.1635	$\pm$	0.3280	&	0.3248	$\pm$	0.0841	&	-1.8051	$\pm$	2.5943	&	AGN	\\ 
KPG492A	&	S0	&	1.6559	$\pm$	0.0812	&	0.4975	$\pm$	-	&	0.5138	$\pm$	0.0994	&	-0.3183	$\pm$	0.1803	&	S-L	\\ 

\enddata \tablenotetext{a}{\tiny{M.T.- Morphological Type. }}
\tablenotetext{.}{\tiny {\textbf {AGN classification denotes those
galaxies that are AGN according to the [N {\scriptsize II}] diagrams but not to the
[S {\scriptsize II}] and/or [O {\scriptsize I}].}}}  \tablenotetext{\dag}{\tiny{\textbf {L-S
classification means that galaxies fall in the separation line for
Seyfert and LINER according to [S {\scriptsize II}] and [O {\scriptsize I}] diagrams.}}}
\tablenotetext{*}{\tiny{\textbf {Type with weak broad component in
permitted lines. Seyfert quiantitative classification according to
\citet{1992MNRAS.257..677W}.}}}

\end{deluxetable}

\clearpage

\begin{deluxetable}{ccccccc}
\tabletypesize{\footnotesize}
\label{a2}
\tablecolumns{7}
\tablewidth{0pt}
\tablecaption{Logarithm intensities ratios and their errors for (E+S) sample.}
\tablehead{
    \colhead{Object}			 & 
    \colhead{M.T.\,\tablenotemark{a}}  	 & 
    \colhead{LOG([N {\scriptsize II}]/H$\alpha$)}      & 
    \colhead{LOG([O {\scriptsize III}]/H$\beta$)} 	 &
    \colhead{LOG([S {\scriptsize II}]/H$\alpha$)}      &
    \colhead{LOG([O {\scriptsize I}]/H$\alpha$) }      &
    \colhead{Type }	 }
\startdata
\sidehead{Eliptical}
\hline
KPG055A	& E & 0.4993 $\pm$ 0.1027 & 1.0842 $\pm$ 0.8013 & 0.4597 $\pm$ 0.1329 & -0.9006 $\pm$ 0.9222& S-L \\
KPG089B & E & 0.3029 $\pm$ 0.0896 & 0.7262 $\pm$ 0.4881 & 0.0451 $\pm$ 0.1716 & -0.4415 $\pm$ 0.2174& S-L \\
KPG091B & E & 0.4149 $\pm$ 0.0881 & 0.1889 $\pm$ 0.1558 & 0.3686 $\pm$ 0.1197 & -1.0699 $\pm$ 1.2131& AGN \\
KPG144B & E & -0.4333 $\pm$ 0.0086 & -0.7456 $\pm$ 0.0294& -0.5566 $\pm$ 0.0130 & -1.8652 $\pm$ 0.0627& Star-forming \\
KPG155A & E & -0.6479 $\pm$ 0.0128 & 0.0830 $\pm$ 0.0188 & -0.4416 $\pm$ 0.0158 & -1.3675 $\pm$ 0.0876& Star-forming \\
KPG167A & E & 0.3963 $\pm$ 0.0562 & 0.2965 $\pm$ 0.1364 & 0.2132 $\pm$ 0.0893 & -0.7086 $\pm$ 0.2628& LINER \\
KPG197B & E & 0.1071 $\pm$ 0.1254 & 0.5876 $\pm$ 0.5488 & -0.0212 $\pm$ 0.2225 & -1.3599 $\pm$ 1.8493& AGN \\
KPG198B & E & 0.2118 $\pm$ 0.0666 & 0.4295 $\pm$ 0.1479 & 0.1489$\pm$ 0.0988 & -1.0428 $\pm$ 0.73956& AGN \\
KPG244A & E & 0.2472 $\pm$ 0.1071 & 0.9321 $\pm$ 0.5472 & 0.0682 $\pm$ 0.1872 & -0.3952 $\pm$ 0.2798& S-L \\
KPG254B & E & 0.1830 $\pm$ 0.1103 & 0.7963 $\pm$ 0.6008 & 0.1519 $\pm$ 0.1599 & -0.5364 $\pm$ 0.3098& S-L \\
KPG260B & E & -0.0098 $\pm$ 0.0491 & 0.5369 $\pm$ 0.1214 & -0.0516 $\pm$ 0.0798 & -0.7262 $\pm$ 0.1743& S-L \\
KPG339B & E & 0.3400 $\pm$ 0.1271 & 0.4900 $\pm$ 0.2531 & -0.4000 $\pm$ 0.5529 & -0.6500 $\pm$ 0.7326& S-L \\
KPG363A & E & -0.1500 $\pm$ 0.0048 & 0.3900 $\pm$ 0.0100 & -0.2062 $\pm$ 0.0053 & -1.1900 $\pm$ 0.0286& Sy2 \\
KPG383A & E & -0.0589 $\pm$ 0.0253 & 0.1041 $\pm$ 0.0549 & 0.0646 $\pm$ 0.0323 & -0.6045 $\pm$ 0.0432& LINER \\
KPG394A & E & -0.5482 $\pm$ 0.0103 & -0.3162 $\pm$ 0.0199 & -0.4843 $\pm$ 0.0135 & -1.7759 $\pm$ 0.1200& Star-forming \\
KPG412B & E & 0.0100 $\pm$ 0.0861 & 0.3000 $\pm$ 0.5584 & -0.3400 $\pm$ 0.2379 & -0.5900 $\pm$ 0.1510& AGN \\
KPG414B & E & -1.0658 $\pm$ 0.0064 & 0.5854 $\pm$ 0.0050 & -0.8526 $\pm$ 0.0075	& -1.9263 $\pm$	0.3964& Star-forming\\
KPG432A & E/S0 & 0.1500 $\pm$ 0.1628 & 0.6500 $\pm$ 0.2518 & -0.0800 $\pm$ 0.3282 & -0.79 $\pm$ 1.3913& AGN \\
KPG466B & E/S0 & -0.2432 $\pm$ 0.0054 & -0.5659 $\pm$ 0.0156 & -0.5076 $\pm$ 0.0082 & -1.4698 $\pm$ 0.0223& AGN \\
KPG494B & E & -0.5027 $\pm$ 0.0065 & 0.0234 $\pm$ 0.0087 & -0.6204 $\pm$ 0.0094 & -1.7156 $\pm$ 0.1363& Star-forming \\
KPG565A & E & 0.0575 $\pm$ 0.1217 & 0.3907$\pm$ 0.1637 & 0.1120 $\pm$ 0.1654 & -1.0119 $\pm$ 0.6625& S-L \\

\sidehead{Spheroidal}
\hline
KPG006A & S0 & -0.4195 $\pm$ 0.0048 & -0.7951 $\pm$ 0.0234 & -0.6294 $\pm$ 0.0084 & -1.7974 $\pm$ 0.0364& Star-forming\\
KPG062A & S0/a & 0.2940 $\pm$ 0.0327 & 0.6557 $\pm$ 0.1096 & 0.1072 $\pm$ 0.0540 & -0.4205 $\pm$ 0.0758& LINER\\
KPG162A	& S0 &	0.0361 $\pm$ 0.0650 & 0.3679 $\pm$ 0.1335 & 0.1669 $\pm$ 0.0810	& -0.5131 $\pm$	0.1451& LINER\\
KPG231A	& S0 &	-0.1258 $\pm$	0.0148 & -0.1432 $\pm$	0.0520 & -0.3094 $\pm$	0.0275 & -1.1732 $\pm$	0.0599& AGN\\
KPG243B	& S0 &	-0.2481 $\pm$ 0.0124 &	-0.2337 $\pm$ 0.0358 &	-0.4167 $\pm$ 0.0220 &	-1.2459 $\pm$ 0.0583& AGN\\
KPG269A	& S0 &	-0.3083 $\pm$ 0.0240 &	-0.0619 $\pm$ 0.0434 &	-0.1174 $\pm$ 0.0289 &	-0.6155 $\pm$ 0.0305& LINER\\
KPG275B	& S0 &	-0.3729 $\pm$	0.0071 & -0.6161 $\pm$	0.0192 & -0.5494 $\pm$	0.0106 & -1.7735 $\pm$	0.0488& Star-forming\\
KPG304A	& S0 &	-0.0882 $\pm$ 	0.0262 & 0.0342 $\pm$	0.1529 & -0.2112 $\pm$	0.0514 & -1.5169 $\pm$	0.3948& AGN\\
KPG317B	& S0 &	-0.1411 $\pm$	0.0318 & -0.0846 $\pm$	0.0741 & -0.1459 $\pm$	0.0495 & -1.1398 $\pm$	0.1709& AGN\\
KPG345B	& S0/a & 0.1580 $\pm$	0.0438 & 0.4055 $\pm$	0.2242 & -0.1401 $\pm$	0.0953 & -1.0692 $\pm$	0.2704& S-L\\
KPG374A	& S0 &	0.2022 $\pm$	0.0470 & 0.8882 $\pm$	0.1197 & 0.2350 $\pm$	0.0630 & -0.5566 $\pm$	0.2465& S-L\\
KPG392A	& S0 &	0.2696 $\pm$	0.0880 & 1.0135 $\pm$	0.6311 & 0.1277 $\pm$	0.1462 & -0.6750 $\pm$	0.4153& S-L\\
KPG402B	& S0 &	-0.0239 $\pm$	0.0353 & 0.3299 $\pm$	0.0950 & -0.0300 $\pm$	0.0542 & -1.2199 $\pm$	0.3090& AGN\\
KPG408A	& S0 &	 0.0200	$\pm$   0.0732 &  0.6000   $\pm$   -0.2100  & -0.2200   $\pm$  0.1651 & -0.9400   $\pm$ 0.4780& S-L\\
KPG436A	& S0 &	-0.4464 $\pm$	0.0115 & -0.8387 $\pm$	0.0711 & -0.6157 $\pm$	0.0222 & -0.9604 $\pm$	0.2130& Star-forming\\

\sidehead{Sa}
\hline
KPG055B	&	Sa	&	-0.4164	$\pm$	0.0355	&	-0.7504	$\pm$	0.2962	&	-0.8892	$\pm$	0.1335	&	-1.6462	$\pm$	0.3224& Star-forming \\
KPG188A	&	Sa	&	-0.3864	$\pm$	0.0080	&	-0.6397	$\pm$	0.0235	&	-0.5991	$\pm$	0.0130	&	-2.0507	$\pm$	0.1187& Star-forming\\
KPG197A	&	Sa	&	0.3868	$\pm$	0.1200	&	0.0900	$\pm$	0.1292	&	0.3765	$\pm$	0.1596	&	-0.8110	$\pm$	1.1059& AGN\\
KPG199B	&	Sa	&	-0.3736	$\pm$	0.0231	&	-0.4916	$\pm$	0.1184	&	-0.4074	$\pm$	0.0370	&	-1.6238	$\pm$	0.2468& Star-forming\\
KPG239B	&	Sa	&	0.2471	$\pm$	0.0812	&	-0.0440	$\pm$	0.1153	&	0.1137	$\pm$	0.1344	&	-0.6385	$\pm$	0.3831& LINER\\
KPG260A	&	Sa	&	0.2989	$\pm$	0.0445	&	0.2197	$\pm$	0.0947	&	0.2112	$\pm$	0.0657	&	-0.6162	$\pm$	0.1726& LINER\\
KPG284A	&	Sa	&	-0.0451	$\pm$	0.0223	&	-0.0490	$\pm$	0.0966	&	-0.2627	$\pm$	0.0448	&	-1.1747	$\pm$	0.1157& AGN\\
KPG303B	&       Sa	&       0.2100	$\pm$   0.1207	&       0.7600	$\pm$	-0.6208	&       -0.2500	$\pm$   0.3488	&       -0.8100	$\pm$   0.6206& Sy2\\
KPG317A	&	Sab	&	0.0313	$\pm$	0.0566	&	0.0050	$\pm$	-0.2062	&	-0.6094	$\pm$	0.2568	&	-0.8377	$\pm$	0.2280& AGN\\
KPG402A	&	Sab	&	-0.5588	$\pm$	0.0195	&	-0.0488	$\pm$	0.0470	&	-0.3656	$\pm$	0.0251	&	-1.3444	$\pm$	0.0987& Star-forming\\
KPG439B	&	Sa	&	-0.2339	$\pm$	0.0068	&	-0.3203	$\pm$	0.0247	&	-0.5447	$\pm$	0.0126	&	-2.0767	$\pm$	0.0270& AGN\\
KPG548A	&	Sa	&	0.1074	$\pm$	0.0315	&	0.3850	$\pm$	0.0636	&	0.1196	$\pm$	0.0437	&	-1.7587	$\pm$	0.0811& AGN\\
KPG466A	&	Sa	&	-0.6200	$\pm$	0.0098	&	0.6000	$\pm$	-0.2100	&	-0.1500	$\pm$	0.0122	&	-0.6300	$\pm$	0.0300& S-L\\

\sidehead{Sb}
\hline
KPG089A	&	Sb	&	-0.4280	$\pm$	0.0063	&	-0.3958	$\pm$	0.0129	&	-0.5371	$\pm$	0.0090	&	-1.7760	$\pm$	0.0579& Star-forming\\
KPG148A	&	Sb	&	-0.7930	$\pm$	0.0076	&	0.1195	$\pm$	0.0094	&	-0.6847	$\pm$	0.0094	&	-1.9934	$\pm$	0.1879& Star-forming	\\
KPG162B	&	Sb	&	-0.2290	$\pm$	0.0278	&	-0.1913	$\pm$	0.0809	&	-0.3094	$\pm$	0.0494	&	-1.7259	$\pm$	0.4955& AGN	\\
KPG202B	&	Sb	&	0.3768	$\pm$	0.0374	&	1.0828	$\pm$	0.2444	&	0.2005	$\pm$	0.0582	&	-0.5370	$\pm$	0.1209& S-L	\\
KPG229B	&	Sb	&	-0.4275	$\pm$	0.0133	&	-0.5682	$\pm$	0.0579	&	-0.6502	$\pm$	0.0274	&	-1.7358	$\pm$	0.1053& Star-forming	\\
KPG243A	&	Sb	&	-0.2517	$\pm$	0.0125	&	0.9063	$\pm$	0.0179	&	-0.3040	$\pm$	0.0191	&	-1.2366	$\pm$	0.2450& Sy2	\\
KPG248B	&	Sb	&	-0.0132	$\pm$	0.0176	&	0.2765	$\pm$	0.0935	&	-0.2094	$\pm$	0.0349	&	-1.0981	$\pm$	0.1070& S-L	\\
KPG254A	&	Sbab	&	0.1189	$\pm$	0.0062	&	1.0848	$\pm$	0.0092	&	-0.2198	$\pm$	0.0098	&	-0.7999	$\pm$	0.0984& Sy2	\\
KPG269B	&	Sb	&	-0.5115	$\pm$	0.0073	&	-0.3156	$\pm$	0.0140	&	-0.5588	$\pm$	0.0101	&	-1.7537	$\pm$	0.0710& Star-forming	\\
KPG339A	&	Sb	&	-0.5009	$\pm$	0.0144	&	-0.9664	$\pm$	0.1302	&	-0.5756	$\pm$	0.0244	&	-1.5262	$\pm$	0.0801& Star-forming	\\
KPG345A	&	Sbs(dm)	&	-0.6208	$\pm$	0.0266	&	-0.1099	$\pm$	0.0559	&	-0.2803	$\pm$	0.0271	&	-1.1642	$\pm$	0.1137& Star-forming	\\
KPG374B	&	Sbc	&	-0.4636	$\pm$	0.0162	&	-0.6559	$\pm$	0.1089	&	-0.5556	$\pm$	0.0298	&	-1.6522	$\pm$	0.1230& Star-forming	\\
KPG392B	&	Sb	&	-0.1453	$\pm$	0.0412	&	0.3090	$\pm$	0.0698	&	-0.1000	$\pm$	0.0604	&	-0.9476	$\pm$	0.2024& S-L	\\
KPG408B	&	Sb	&	-0.5318	$\pm$	0.0276	&	-0.0493	$\pm$	0.0680	&	-0.2616	$\pm$	0.0306	&	-1.2143	$\pm$	0.1363& Star-forming	\\
KPG416A	&	Sb	&	-0.4165	$\pm$	0.0079	&	-0.7686	$\pm$	0.0294	&	-0.6609	$\pm$	0.0129	&	-2.0159	$\pm$	0.0828& Star-forming	\\
KPG419A	&	SBb	&	-0.0400	$\pm$	0.0073	&	1.0000	$\pm$	-0.2100	&	-0.1400	$\pm$	0.0082	&	-0.8800	$\pm$	0.0654& Sy1.9	\\
KPG432B	&	Sb	&	-0.5281	$\pm$	0.0183	&	-0.8646	$\pm$	0.1123	&	-0.5979	$\pm$	0.0308	&	-0.8139	$\pm$	0.6375& Star-forming	\\
KPG436B	&	Sb	&	-0.3351	$\pm$	0.0260	&	-0.2282	$\pm$	0.1648	&	-0.5081	$\pm$	0.0554	&	-2.3363	$\pm$	0.1890& Star-forming	\\
KPG484B	&	Sb	&	-0.4621	$\pm$	0.0242	&	-0.5690	$\pm$	0.1357	&	-0.6541	$\pm$	0.0586	&	-1.3153	$\pm$	0.5506& Star-forming	\\
KPG494A	&	Sbc	&	-0.2737	$\pm$	0.0106	&	-0.4524	$\pm$	0.0416	&	-0.4959	$\pm$	0.0207	&	-0.6307	$\pm$	0.1096& AGN	\\
KPG497B	&	Sb	&	0.2149	$\pm$	0.0372	&	0.1282	$\pm$	0.1041	&	0.2410	$\pm$	0.0495	&	-1.9189	$\pm$	0.2746& AGN	\\
KPG558A	&	Sb	&	-0.3820	$\pm$	0.0216	&	-0.5685	$\pm$	0.1345	&	-0.5350	$\pm$	0.0464	&	-1.8762	$\pm$	2.0941& Star-forming	\\
KPG565B	&	Sb	&	-0.5528	$\pm$	0.0163	&	-0.2266	$\pm$	0.0389	&	-0.4436	$\pm$	0.0215	&	-0.5792	$\pm$	0.1180& Star-forming	\\

\\
\sidehead{Sc}
\hline
KPG058A	&	Sc	&	-0.2334	$\pm$	0.0200	&	-0.3081	$\pm$	0.1289	&	-0.3386	$\pm$	0.0395	&	-1.2968	$\pm$	0.1063& AGN	\\
KPG287A	&	Sc	&	-0.4494	$\pm$	0.0290	&	-0.7093	$\pm$	0.2075	&	-0.5882	$\pm$	0.0640	&	-1.7455	$\pm$	0.4297& Star-forming	\\
KPG314B	&	Sc	&	-0.4574	$\pm$	0.0225	&	-0.6110	$\pm$	0.1444	&	-0.2951	$\pm$	0.0290	&	-1.2424	$\pm$	0.0949& Star-forming	\\
KPG353A	&	Sc	&	-0.6103	$\pm$	0.0155	&	-0.9310	$\pm$	0.1019	&	-0.8004	$\pm$	0.0309	&	-2.1018	$\pm$	0.2166& Star-forming	\\
KPG407A	&	Sc	&	-0.1260	$\pm$	0.0093	&	0.1290	$\pm$	0.0184	&	-0.4357	$\pm$	0.0173	&	-1.3858	$\pm$	0.0731& AGN	\\
KPG412A	&	Sc	&	-0.3194	$\pm$	0.0167	&	-0.2700	$\pm$	0.0974	&	-0.4120	$\pm$	0.0315	&	-1.3003	$\pm$	0.0892& Star-forming	\\
KPG464B	&	Sc	&	-0.3634	$\pm$	0.0338	&	-0.6033	$\pm$	0.2548	&	-0.6392	$\pm$	0.0926	&	-1.4960	$\pm$	0.1895& Star-forming	\\

\sidehead{Sm}
\hline
KPG198A	& Sm &	-0.6659 $\pm$ 0.0053 & 0.0858 $\pm$ 0.0066 & -0.6191 $\pm$ 0.0066 & -1.6593 $\pm$ 0.0651& Star-forming\\
KPG414A	& Sm &	-0.7539 $\pm$ 0.0291 &	0.1398 $\pm$ 0.0340 & -0.3100 $\pm$ 0.0247 & -1.4477 $\pm$ 0.2577& Star-forming\\

\enddata

\tablenotetext{a}{\tiny{M.T.- Morphological Type. }}
\tablenotetext{.}{\tiny {\textbf {AGN classification denotes those
galaxies that are AGN according to the [N {\scriptsize II}] diagrams but not to the
[S {\scriptsize II}] and/or [O {\scriptsize I}].}}}  \tablenotetext{\dag}{\tiny{\textbf {L-S
classification means that galaxies fall in the separation line for
Seyfert and LINER according to [S {\scriptsize II}] and [O {\scriptsize I}] diagrams.}}}
\tablenotetext{*}{\tiny{\textbf {Type with weak broad component in
permitted lines. Seyfert quiantitative classification according to
\citet{1992MNRAS.257..677W}.}}}
\end{deluxetable}

\clearpage

\begin{deluxetable}{ccccccc}
\tabletypesize{\footnotesize}
\label{a3}
\tablecolumns{7}
\tablewidth{0pt}
\tablecaption{Logarithm intensities ratios and their errors for (S+S) sample.}
\tablehead{
    \colhead{Object}			 & 
    \colhead{M.T.\,\tablenotemark{a}}  	 & 
    \colhead{LOG([N {\scriptsize II}]/H$\alpha$)}      & 
    \colhead{LOG([O {\scriptsize III}]/H$\beta$)} 	 &
    \colhead{LOG([S {\scriptsize II}]/H$\alpha$)}      &
    \colhead{LOG([O {\scriptsize I}]/H$\alpha$) }      &
    \colhead{Type }	 }
\startdata

\sidehead{Sa}
\hline
KPG052B	&	SBa	&	-0.0227	$\pm$	0.0390	&	0.3490	$\pm$	0.0996	&	0.1075	$\pm$	0.0491	&	-0.7634	$\pm$	0.1161	&	LINER	\\
KPG161A	&	Sa	&	-0.5634	$\pm$	0.0375	&	0.0263	$\pm$	0.0066	&	-0.5587	$\pm$	0.0491	&	-1.6507	$\pm$	0.0301	&	Star-forming	\\
KPG174B	&	Sab	&	0.2549	$\pm$	0.0612	&	0.2078	$\pm$	0.0900	&	0.1152	$\pm$	0.0763	&	-0.8191	$\pm$	0.1435	&	LINER	\\
KPG185B	&	Sab	&	-0.2759	$\pm$	0.0069	&	-0.2778	$\pm$	0.0168	&	-0.5190	$\pm$	0.0130	&	-1.6973	$\pm$	0.0616	&	AGN	\\
KPG190B	&	Sab	&	0.1749	$\pm$	0.0543	&	0.3164	$\pm$	0.1391	&	0.0379	$\pm$	0.0914	&	-0.7517	$\pm$	0.2010	&	LINER	\\
KPG201A	&	Sa	&	0.4490	$\pm$	0.0698	&	1.4683	$\pm$	1.8727	&	0.4423	$\pm$	0.0912	&	-0.3428	$\pm$	0.1627	&	S-L	\\
KPG211A	&	Sa	&	0.4428	$\pm$	0.0748	&	0.7200	$\pm$	0.2252	&	0.1979	$\pm$	0.1218	&	-0.5703	$\pm$	0.1677	&	LINER	\\
KPG213B	&	Sa	&	0.3444	$\pm$	0.0579	&	0.2904	$\pm$	0.1466	&	0.2517	$\pm$	0.0891	&	-0.4891	$\pm$	0.1528	&	LINER	\\
KPG215A	&	Sab	&	-0.1434	$\pm$	0.0328	&	0.0360	$\pm$	0.1589	&	-0.2980	$\pm$	0.0655	&	-1.4753	$\pm$	0.3560	&	LINER	\\
KPG220B	&	SBa	&	-0.1989	$\pm$	0.0099	&	-0.2887	$\pm$	0.0302	&	-0.3935	$\pm$	0.0168	&	-1.3904	$\pm$	0.0502	&	AGN	\\
KPG222A	&	Sa	&	-0.8551	$\pm$	0.0166	&	0.2688	$\pm$	0.0138	&	-0.4264	$\pm$	0.0146	&	-1.3003	$\pm$	0.0321	&	Star-forming	\\
KPG226A	&	Sa	&	-0.3940	$\pm$	0.0285	&	-0.4566	$\pm$	0.0566	&	-0.6650	$\pm$	0.0497	&	-1.7715	$\pm$	0.1479	&	Star-forming	\\
KPG228A	&	Sa	&	0.6609	$\pm$	0.0571	&	0.9678	$\pm$	0.4890	&	0.5192	$\pm$	0.0759	&	-0.5400	$\pm$	0.1937	&	LINER	\\
KPG241B	&	Sab	&	0.2447	$\pm$	0.0812	&	0.2500	$\pm$	0.1807	&	-0.0550	$\pm$	0.1654	&	-0.9701	$\pm$	0.4729	&	LINER	\\
KPG246A	&	Sa	&	-0.2320	$\pm$	0.0315	&	-0.1986	$\pm$	0.1405	&	-0.3219	$\pm$	0.0585	&	-1.0657	$\pm$	0.1296	&	AGN	\\
KPG252A	&	Sa	&	-0.3869	$\pm$	0.0086	&	-0.3767	$\pm$	0.0242	&	-0.4692	$\pm$	0.0130	&	-1.4058	$\pm$	0.0306	&	Star-forming	\\
KPG255A	&	Sab	&	-0.5744	$\pm$	0.0633	&	0.0659	$\pm$	0.1759	&	-0.2047	$\pm$	0.0573	&	-1.1323	$\pm$	0.2016	&	Star-forming	\\
KPG257B	&	Sab	&	-0.3510	$\pm$	0.0424	&	-0.4617	$\pm$	0.1430	&	-0.3507	$\pm$	0.0626	&	-1.3867	$\pm$	0.1455	&	Star-forming	\\
KPG265B	&	Sa	&	-0.5187	$\pm$	0.0063	&	-0.2686	$\pm$	0.0157	&	-0.4742	$\pm$	0.0083	&	-1.7040	$\pm$	0.0328	&	Star-forming	\\
KPG266A	&	Sab	&	0.2197	$\pm$	0.0233	&	0.2318	$\pm$	0.0660	&	0.0383	$\pm$	0.0383	&	-0.6966	$\pm$	0.0718	&	LINER	\\
KPG270A	&	Sab	&	-0.3889	$\pm$	0.0286	&	-0.6248	$\pm$	0.1815	&	-0.5075	$\pm$	0.0526	&	-1.8829	$\pm$	0.5054	&	Star-forming	\\
KPG285A	&	Sab	&	0.1228	$\pm$	0.0315	&	0.3531	$\pm$	0.1212	&	0.0166	$\pm$	0.0511	&	-1.0220	$\pm$	0.1804	&	LINER	\\
KPG289B	&	Sa	&	-0.5191	$\pm$	0.0108	&	-0.8371	$\pm$	0.0614	&	-0.8041	$\pm$	0.0242	&	-1.7303	$\pm$	0.0709	&	Star-forming	\\
KPG312B	&	S0	&	0.0515	$\pm$	0.0162	&	-0.0333	$\pm$	0.0438	&	-0.3137	$\pm$	0.0367	&	-1.0605	$\pm$	0.0770	&	AGN	\\
KPG318A	&	Sa	&	-0.0607	$\pm$	0.0213	&	0.1645	$\pm$	0.0371	&	-0.2907	$\pm$	0.0402	&	-1.3473	$\pm$	0.1531	&	S-L	\\
KPG327A	&	S0/a	&	0.1601	$\pm$	0.0569	&	0.5441	$\pm$	0.1769	&	0.0906	$\pm$	0.0883	&	-0.8800	$\pm$	0.2621	&	LINER	\\
KPG327B	&	S0/a	&	-0.3217	$\pm$	0.0079	&	-0.6363	$\pm$	0.0345	&	-0.5507	$\pm$	0.0141	&	-1.6678	$\pm$	0.0495	&	Star-forming	\\
KPG337A	&	SABa	&	0.0719	$\pm$	0.0000	&	1.3675	$\pm$	1.9245	&	-0.3788	$\pm$	0.0559	&	-0.7109	$\pm$	0.0000	&	Sy2	\\
KPG342A	&	Sa	&	-0.5285	$\pm$	0.0085	&	-0.3034	$\pm$	0.0160	&	-0.5119	$\pm$	0.0118	&	-1.8031	$\pm$	0.0491	&	Star-forming	\\
KPG348A	&	Sa	& 	-0.1799	$\pm$	0.0124	&	0.0827	$\pm$	0.0168	&	-0.1113	$\pm$	0.0146	&	-1.0213	$\pm$	0.0319	&	AGN*	\\
KPG349B	&	SABm	&	-0.2769	$\pm$	0.0287	&	-0.3732	$\pm$	0.1181	&	-0.2745	$\pm$	0.0429	&	-1.0415	$\pm$	0.0947	&	AGN	\\
KPG355A	&	S0/a	&	-0.2799	$\pm$	0.0102	&	-0.0799	$\pm$	0.0338	&	-0.3937	$\pm$	0.0173	&	-1.5822	$\pm$	0.0717	&	AGN	\\
KPG360A	&	Sab	&	0.1978	$\pm$	0.0842	&	0.6304	$\pm$	0.1412	&	0.2513	$\pm$	0.1100	&	-0.4534	$\pm$	0.1876	&	LINER	\\
KPG366A	&	S0/a	&	-0.3838	$\pm$	0.0145	&	-0.5970	$\pm$	0.0743	&	-0.5617	$\pm$	0.0291	&	-1.7457	$\pm$	0.1496	&	Star-forming	\\
KPG368A	&	Sa	&	-0.0668	$\pm$	0.0168	&	0.7033	$\pm$	0.0284	&	-0.2514	$\pm$	0.0306	&	-0.9945	$\pm$	0.0627	&	Sy2	\\
KPG369B	&	Sab	&	-0.3745	$\pm$	0.0043	&	0.0370	$\pm$	0.0046	&	-0.6268	$\pm$	0.0061	&	-1.5678	$\pm$	0.0101	&	AGN	\\
KPG384A	&	SABd	&	-1.2366	$\pm$	0.0395	&	0.4807	$\pm$	0.0155	&	-0.6179	$\pm$	0.0234	&	-1.8310	$\pm$	0.1430	&	Star-forming	\\
KPG385A	&	Sab	&	-0.2753	$\pm$	0.0144	&	-0.1326	$\pm$	0.0389	&	-0.3423	$\pm$	0.0224	&	-1.0884	$\pm$	0.0442	&	AGN*	\\
KPG395B	&	Sa	&	-0.4141	$\pm$	0.0192	&	-0.5530	$\pm$	0.1063	&	-0.5156	$\pm$	0.0351	&	-1.7811	$\pm$	0.2259	&	Star-forming	\\
KPG400B	&	Sa	&	-0.4082	$\pm$	0.0112	&	-0.7861	$\pm$	0.0539	&	-0.5737	$\pm$	0.0192	&	-1.8396	$\pm$	0.1003	&	Star-forming	\\
KPG413A	&	Sa	&	0.1181	$\pm$	0.0441	&	0.3457	$\pm$	0.0734	&	0.1669	$\pm$	0.0605	&	-0.7199	$\pm$	0.1441	&	LINER	\\
KPG423A	&	Sab	&	-0.2772	$\pm$	0.0156	&	-0.3584	$\pm$	0.0750	&	-0.6121	$\pm$	0.0392	&	-1.4676	$\pm$	0.1124	&	AGN	\\
KPG425A	&	SBa	&	-0.4318	$\pm$	0.0080	&	-0.5628	$\pm$	0.0166	&	-0.5251	$\pm$	0.0114	&	-1.8760	$\pm$	0.0501	&	Star-forming	\\
KPG425B	&	S0/a	&	-0.0411	$\pm$	0.0260	&	0.8092	$\pm$	0.0597	&	-0.1732	$\pm$	0.0475	&	-1.2296	$\pm$	0.1709	&	Sy2	\\
KPG437B	&	Sab	&	-0.1406	$\pm$	0.0056	&	0.8404	$\pm$	0.0079	&	-0.5234	$\pm$	0.0099	&	-1.3326	$\pm$	0.0184	&	Sy2	\\
KPG447A	&	Sa	&	-0.3622	$\pm$	0.0120	&	-0.4357	$\pm$	0.0390	&	-0.4910	$\pm$	0.0198	&	-1.7377	$\pm$	0.0999	&	Star-forming	\\
KPG449B	&	Sab	&	-0.1371	$\pm$	0.0189	&	-0.1516	$\pm$	0.0570	&	-0.2620	$\pm$	0.0316	&	-1.1676	$\pm$	0.0897	&	AGN	\\
KPG455A	&	Sa	&	0.3636	$\pm$	0.0750	&	0.9361	$\pm$	59.0739	&	0.1912	$\pm$	0.1196	&	-0.7281	$\pm$	0.3089	&	LINER	\\
KPG474B	&	Sa	&	-0.5031	$\pm$	0.0114	&	-0.7489	$\pm$	0.0666	&	-0.6821	$\pm$	0.0237	&	-1.6340	$\pm$	0.0783	&	Star-forming	\\
KPG477B	&	S0	&	0.2623	$\pm$	0.0419	&	0.4125	$\pm$	0.1388	&	0.1872	$\pm$	0.0615	&	-0.6152	$\pm$	0.1262	&	LINER	\\
KPG480B	&	S0a	&	0.3424	$\pm$	0.0285	&	0.9104	$\pm$	0.1547	&	0.0440	$\pm$	0.0531	&	-0.4339	$\pm$	0.0653	&	S-L	\\
KPG495B	&	SBab	&	-0.3551	$\pm$	0.0309	&	-0.9646	$\pm$	0.5063	&	-0.5157	$\pm$	0.0663	&	-1.3328	$\pm$	0.1772	&	Star-forming	\\
KPG496A	&	S0	&	-0.1891	$\pm$	0.0222	&	0.0000	$\pm$	0.0976	&	-0.4839	$\pm$	0.0572	&	-1.4065	$\pm$	0.1872	&	AGN	\\
KPG150A	&	Sa 	&	0.3299	$\pm$	0.0499	&	0.9581	$\pm$	0.1379	&	0.1766	$\pm$	0.0791	&	-0.6516	$\pm$	0.1713	&	LINER	\\
KPG286B	&	Sa 	&	0.6966	$\pm$	0.2316	&	0.0401	$\pm$	0.0193	&	-0.3559	$\pm$	0.0159	&	-1.1296	$\pm$	0.0055	&	AGN	\\
KPG355B	&	Sa 	&	-0.1845	$\pm$	0.0111	&	-0.1347	$\pm$	0.0255	&	0.0186	$\pm$	0.0130	&	-0.6826	$\pm$	0.0209	&	LINER	\\
KPG368B	&	Sa 	&	-0.2844	$\pm$	0.0063	&	0.3010	$\pm$	0.0084	&	-0.1928	$\pm$	0.0078	&	-0.8456	$\pm$	0.0102	&	LINER	\\
KPG378B	&	SABb	&	0.2798	$\pm$	0.0172	&	0.4255	$\pm$	0.0470	&	0.1171	$\pm$	0.0268	&	-0.5858	$\pm$	0.0457	&	LINER	\\
KPG388B	&	Sab 	&	-0.2007	$\pm$	0.0049	&	0.0933	$\pm$	0.0058	&	-0.3396	$\pm$	0.0065	&	-0.9527	$\pm$	0.0077	&	AGN	\\
KPG398A	&	Sab 	&	-0.4541	$\pm$	0.0056	&	0.9582	$\pm$	0.0055	&	-0.7017	$\pm$	0.0102	&	-0.9831	$\pm$	0.0096	&	Sy2	\\

\sidehead{Sb}
\hline
KPG022B	&	Sbb	&	-0.1212	$\pm$	0.0147	&	-0.1473	$\pm$	0.0777	&	-0.3904	$\pm$	0.0314	&	-1.4088	$\pm$	0.1171	&	AGN	\\
KPG049A	&	Sb	&	0.1339	$\pm$	0.0282	&	0.4632	$\pm$	0.0831	&	-0.1497	$\pm$	0.0587	&	-0.8720	$\pm$	0.1155	&	S-L	\\
KPG053B	&	Sbb	&	-0.1190	$\pm$	0.0520	&	0.6112	$\pm$	0.1749	&	-0.2358	$\pm$	0.1017	&	-1.6885	$\pm$	0.9768	&	Sy2	\\
KPG146A	&	Sb	&	-0.3627	$\pm$	0.0189	&	-0.2041	$\pm$	0.0486	&	-0.5772	$\pm$	0.0351	&	-1.4183	$\pm$	0.0865	&	AGN	\\
KPG150B	&	Sb	&	0.2416	$\pm$	0.0323	&	0.2774	$\pm$	0.0938	&	0.1807	$\pm$	0.0462	&	-0.6013	$\pm$	0.0887	&	LINER	\\
KPG159B	&	Sb	&	-0.3331	$\pm$	0.0310	&	-0.1897	$\pm$	0.0773	&	-0.3514	$\pm$	0.0512	&	-1.4715	$\pm$	0.2742	&	Star-forming	\\
KPG163A	&	Sbc	&	-0.4105	$\pm$	0.0292	&	-0.6588	$\pm$	0.2047	&	-0.5693	$\pm$	0.0666	&	-1.4852	$\pm$	0.2231	&	Star-forming	\\
KPG168B	&	Sc	&	-0.6883	$\pm$	0.0060	&	-0.0187	$\pm$	0.0075	&	-0.8003	$\pm$	0.0078	&	-2.2243	$\pm$	0.0307	&	Star-forming	\\
KPG171A	&	Sbc	&	-0.7311	$\pm$	0.0115	&	0.2867	$\pm$	0.0141	&	-0.5201	$\pm$	0.0135	&	-1.4738	$\pm$	0.0332	&	Star-forming	\\
KPG171B	&	Sb	&	-0.4419	$\pm$	0.0085	&	-0.3946	$\pm$	0.0230	&	-0.5073	$\pm$	0.0125	&	-1.6458	$\pm$	0.0451	&	Star-forming	\\
KPG174A	&	Sbc	&	-0.5573	$\pm$	0.0201	&	-0.3020	$\pm$	0.0520	&	-0.3529	$\pm$	0.0257	&	-1.3068	$\pm$	0.0740	&	Star-forming	\\
KPG178B	&	Sb	&	-0.7183	$\pm$	0.0142	&	0.3040	$\pm$	0.0150	&	-0.4359	$\pm$	0.0157	&	-1.4182	$\pm$	0.0422	&	Star-forming	\\
KPG179B	&	Sb	&	-0.0412	$\pm$	0.0464	&	0.2803	$\pm$	0.0994	&	0.0372	$\pm$	0.0623	&	-0.8445	$\pm$	0.1574	&	S-L	\\
KPG193B	&	Sb	&	-0.1522	$\pm$	0.0152	&	0.0000	$\pm$	0.0936	&	-0.3304	$\pm$	0.0288	&	-1.3312	$\pm$	0.0911	&	AGN	\\
KPG200B	&	Sb	&	-0.4491	$\pm$	0.0052	&	-0.1530	$\pm$	0.0080	&	-0.5813	$\pm$	0.0072	&	-1.8146	$\pm$	0.0219	&	Star-forming	\\
KPG205B	&	Sb	&	-0.3443	$\pm$	0.0296	&	-0.4197	$\pm$	0.1369	&	-0.5707	$\pm$	0.0555	&	-1.5058	$\pm$	0.0627	&	Star-forming	\\
KPG206A	&	Sb	&	-0.8756	$\pm$	0.0178	&	0.2553	$\pm$	0.0124	&	-0.4909	$\pm$	0.0151	&	-1.4872	$\pm$	0.0417	&	Star-forming	\\
KPG215B	&	Sb	&	-0.2622	$\pm$	0.0239	&	0.0711	$\pm$	0.0420	&	-0.5030	$\pm$	0.0453	&	-1.4455	$\pm$	0.0635	&	AGN	\\
KPG216B	&	Sbc	&	-0.6998	$\pm$	0.0110	&	0.1614	$\pm$	0.0152	&	-0.4748	$\pm$	0.0131	&	-1.5286	$\pm$	0.0489	&	Star-forming	\\
KPG219B	&	Sb	&	-0.3349	$\pm$	0.0352	&	-0.6951	$\pm$	0.1467	&	-0.5481	$\pm$	0.0619	&	-1.6193	$\pm$	0.0652	&	Star-forming	\\
KPG225A	&	Sb	&	-0.3561	$\pm$	0.0392	&	-0.4868	$\pm$	0.0688	&	-0.3852	$\pm$	0.0491	&	-1.2769	$\pm$	0.0931	&	Star-forming	\\
KPG225B	&	Sb	&	-0.2276	$\pm$	0.0130	&	-0.3010	$\pm$	0.0614	&	-0.4369	$\pm$	0.0258	&	-1.2730	$\pm$	0.0656	&	AGN	\\
KPG226B	&	Sb	&	0.2264	$\pm$	0.0506	&	0.0969	$\pm$	0.0844	&	0.2264	$\pm$	0.0690	&	-0.5540	$\pm$	0.1575	&	LINER	\\
KPG227B	&	Sb	&	-0.6104	$\pm$	0.0110	&	-0.0578	$\pm$	0.0175	&	-0.3952	$\pm$	0.0128	&	-1.6618	$\pm$	0.0590	&	Star-forming	\\
KPG240A	&	Sbc	&	0.2991	$\pm$	0.5267	&	0.9284	$\pm$	0.1415	&	-0.2885	$\pm$	0.0649	&	-0.5655	$\pm$	0.1300	&	Sy2	\\
KPG242A	&	Sb	&	-1.1193	$\pm$	0.1114	&	0.4814	$\pm$	0.0118	&	-0.6282	$\pm$	0.0363	&	-1.5874	$\pm$	0.0727	&	Star-forming	\\
KPG245B	&	Sb	&	0.1627	$\pm$	0.0400	&	0.3357	$\pm$	0.1047	&	0.0205	$\pm$	0.0660	&	-1.0810	$\pm$	0.2713	&	LINER	\\
KPG257A	&	Sb	&	-0.4876	$\pm$	0.0266	&	-0.2314	$\pm$	0.0898	&	-0.2550	$\pm$	0.0296	&	-1.2567	$\pm$	0.1024	&	Star-forming	\\
KPG261A	&	Sb	&	-0.3536	$\pm$	0.0225	&	-0.4393	$\pm$	0.0897	&	-0.4580	$\pm$	0.0416	&	-1.2600	$\pm$	0.1160	&	Star-forming	\\
KPG261B	&	Sb	&	-0.4749	$\pm$	0.0354	&	-0.7247	$\pm$	0.1746	&	-0.6270	$\pm$	0.0707	&	-1.5737	$\pm$	0.1622	&	Star-forming	\\
KPG266B	&	Sb	&	-0.5208	$\pm$	0.0195	&	-0.2194	$\pm$	0.0601	&	-0.3808	$\pm$	0.0269	&	-1.5672	$\pm$	0.1464	&	Star-forming	\\
KPG270B	&	Sb	&	0.2372	$\pm$	0.0993	&	0.1943	$\pm$	0.1247	&	0.2138	$\pm$	0.1373	&	-0.4133	$\pm$	0.2153	&	LINER	\\
KPG283B	&	Sb	&	-1.1220	$\pm$	0.0158	&	0.5031	$\pm$	0.0121	&	-0.6785	$\pm$	0.0139	&	-1.7420	$\pm$	0.0507	&	Star-forming	\\
KPG289A	&	Sb	&	-0.2022	$\pm$	0.0161	&	-0.0746	$\pm$	0.0431	&	-0.3911	$\pm$	0.0313	&	-1.3118	$\pm$	0.0922	&	AGN	\\
KPG291A	&	Sbc	&	-0.2478	$\pm$	0.0183	&	-0.0911	$\pm$	0.0752	&	-0.3840	$\pm$	0.0363	&	-1.8866	$\pm$	0.4021	&	AGN	\\
KPG293A	&	Sb	&	-0.3586	$\pm$	0.0095	&	-0.3903	$\pm$	0.0327	&	-0.4269	$\pm$	0.0153	&	-1.5016	$\pm$	0.0569	&	Star-forming	\\
KPG294A	&	Sb	&	-0.3972	$\pm$	0.0191	&	-0.2596	$\pm$	0.0482	&	-0.2512	$\pm$	0.0235	&	-1.4523	$\pm$	0.1082	&	Star-forming	\\
KPG296A	&	SBb	&	-0.0555	$\pm$	0.0299	&	0.2300	$\pm$	0.0682	&	-0.1711	$\pm$	0.0515	&	-1.0046	$\pm$	0.1286	&	LINER	\\
KPG298A	&	Sbc	&	-0.1960	$\pm$	0.0399	&	-0.3293	$\pm$	0.1168	&	-0.4070	$\pm$	0.0917	&	-1.4421	$\pm$	0.3712	&	AGN	\\
KPG298B	&	Sbc	&	0.1259	$\pm$	0.0537	&	0.8943	$\pm$	0.4159	&	0.0713	$\pm$	0.0834	&	-0.9495	$\pm$	0.2633	&	S-L	\\
KPG299A	&	Sb	&	0.3614	$\pm$	0.0275	&	0.6642	$\pm$	0.0974	&	0.3055	$\pm$	0.0377	&	-0.5099	$\pm$	0.0716	&	LINER	\\
KPG301A	&	Sb	&	-0.4259	$\pm$	0.0101	&	-0.8002	$\pm$	0.0459	&	-0.7081	$\pm$	0.0211	&	-1.6543	$\pm$	0.0636	&	Star-forming	\\
KPG306B	&	Sb	&	-0.2486	$\pm$	0.0073	&	-0.8852	$\pm$	0.0379	&	-0.6573	$\pm$	0.0147	&	-2.0341	$\pm$	0.0968	&	Star-forming	\\
KPG307A	&	Sb	&	-0.4932	$\pm$	0.0064	&	-0.2756	$\pm$	0.0094	&	-0.6434	$\pm$	0.0099	&	-1.9298	$\pm$	0.0341	&	Star-forming	\\
KPG313A	&	Sbc	&	-0.4405	$\pm$	0.0205	&	-0.5497	$\pm$	0.0759	&	-0.3869	$\pm$	0.0283	&	-1.5895	$\pm$	0.1532	&	Star-forming	\\
KPG313B	&	Sb	&	-0.0556	$\pm$	0.0158	&	0.6717	$\pm$	0.0276	&	-0.3918	$\pm$	0.0340	&	-1.2583	$\pm$	0.0855	&	Sy2	\\
KPG319A	&	Sb	&	-0.9226	$\pm$	0.0147	&	0.5866	$\pm$	0.0120	&	-0.6614	$\pm$	0.0167	&	-1.6406	$\pm$	0.0467	&	Star-forming	\\
KPG322B	&	Sbc	&	-0.2800	$\pm$	0.0038	&	-0.6701	$\pm$	0.0083	&	-0.6007	$\pm$	0.0056	&	-1.8646	$\pm$	0.0157	&	Star-forming	\\
KPG323A	&	SBbc	&	-0.5878	$\pm$	0.0179	&	-0.2800	$\pm$	0.0410	&	-0.3375	$\pm$	0.0221	&	-1.4370	$\pm$	0.0741	&	Star-forming	\\
KPG323B	&	Sbc	&	-0.4803	$\pm$	0.0239	&	-0.2004	$\pm$	0.0598	&	-0.3376	$\pm$	0.0336	&	-1.2207	$\pm$	0.0834	&	Star-forming	\\
KPG329B	&	Sb	&	-0.3388	$\pm$	0.0082	&	-0.8493	$\pm$	0.0334	&	-0.6378	$\pm$	0.0147	&	-1.8235	$\pm$	0.1893	&	Star-forming	\\
KPG337B	&	Sb	&	-1.0264	$\pm$	0.0152	&	0.3467	$\pm$	0.0126	&	-0.6109	$\pm$	0.0139	&	-1.7134	$\pm$	0.0522	&	Star-forming	\\
KPG347A	&	Sbc	&	-0.5044	$\pm$	0.0196	&	-1.1862	$\pm$	0.3571	&	-0.7406	$\pm$	0.0449	&	-1.9546	$\pm$	0.2896	&	Star-forming	\\
KPG347B	&	Sbc	&	-0.3895	$\pm$	0.0136	&	-0.4110	$\pm$	0.0496	&	-0.7557	$\pm$	0.0338	&	-1.9613	$\pm$	0.1864	&	Star-forming	\\
KPG352A	&	SBb	&	0.0507	$\pm$	0.0545	&	0.2041	$\pm$	0.1114	&	-0.2545	$\pm$	0.1282	&	-1.7096	$\pm$	1.3149	&	S-L	\\
KPG360B	&	Sb	&	-0.2248	$\pm$	0.0376	&	-0.3919	$\pm$	0.1875	&	-0.3567	$\pm$	0.0812	&	-1.1835	$\pm$	0.1990	&	Star-forming	\\
KPG366B	&	SBb	&	0.3802	$\pm$	0.0669	&	0.5819	$\pm$	0.1983	&	0.2094	$\pm$	0.1066	&	-0.4882	$\pm$	0.1790	&	LINER	\\
KPG375A	&	SBbc	&	-0.2547	$\pm$	0.0074	&	-0.4952	$\pm$	0.0257	&	-0.6052	$\pm$	0.0140	&	-1.8573	$\pm$	0.0695	&	AGN	\\
KPG378A	&	Sb	&	0.0231	$\pm$	0.0197	&	-0.0637	$\pm$	0.0670	&	-0.1819	$\pm$	0.0375	&	-1.2019	$\pm$	0.1338	&	LINER	\\
KPG381B	& 	Sbc	&	-0.2224	$\pm$	0.0122	&	0.2234	$\pm$	0.0205	&	-0.3702	$\pm$	0.0187	&	-1.1871	$\pm$	0.0390	&	Sy1.8	\\
KPG382A	&	Sbc	&	-0.5246	$\pm$	0.0182	&	-0.6493	$\pm$	0.0810	&	-0.3999	$\pm$	0.0268	&	-1.3977	$\pm$	0.0793	&	Star-forming	\\
KPG395A	&	Sbc	&	-0.4479	$\pm$	0.0164	&	-0.7495	$\pm$	0.1079	&	-0.5490	$\pm$	0.0293	&	-1.6905	$\pm$	0.1385	&	Star-forming	\\
KPG400A	&	Sb	&	-0.1371	$\pm$	0.0348	&	-0.0136	$\pm$	0.1297	&	-0.5150	$\pm$	0.1049	&	-1.1601	$\pm$	0.1908	&	AGN	\\
KPG403B	&	Sb	&	-0.4262	$\pm$	0.0124	&	-0.3931	$\pm$	0.0558	&	-0.4494	$\pm$	0.0202	&	-1.7366	$\pm$	0.1136	&	Star-forming	\\
KPG404B	&	Sb	&	0.2027	$\pm$	0.0368	&	0.3643	$\pm$	0.0746	&	0.0242	$\pm$	0.0627	&	-0.7755	$\pm$	0.1349	&	LINER	\\
KPG405A	&	Sb	&	-0.5513	$\pm$	0.0106	&	-0.2386	$\pm$	0.0225	&	-0.4504	$\pm$	0.0138	&	-1.6452	$\pm$	0.0503	&	Star-forming	\\
KPG406A	&	Sb	&	-0.5073	$\pm$	0.0170	&	-0.2486	$\pm$	0.0390	&	-0.3416	$\pm$	0.0209	&	-1.5003	$\pm$	0.0891	&	Star-forming	\\
KPG413B	&	Sbc	&	-0.3930	$\pm$	0.0088	&	-0.8571	$\pm$	0.0582	&	-0.6651	$\pm$	0.0175	&	-1.8924	$\pm$	0.0907	&	Star-forming	\\
KPG426B	&	Sb	&	-0.3316	$\pm$	0.0105	&	-0.6744	$\pm$	0.0464	&	-0.5643	$\pm$	0.0196	&	-1.7752	$\pm$	0.0979	&	Star-forming	\\
KPG427B	&	Sb	&	-0.1189	$\pm$	0.0000	&	-0.0786	$\pm$	0.0981	&	-0.5960	$\pm$	0.1585	&	-1.4024	$\pm$	0.0000	&	AGN	\\
KPG428A	&	SBb	&	-0.7162	$\pm$	0.0286	&	0.0823	$\pm$	0.0445	&	-0.2671	$\pm$	0.0247	&	-1.3747	$\pm$	0.0990	&	Star-forming	\\
KPG430B	&	Sbc	&	-0.6538	$\pm$	0.0130	&	0.1317	$\pm$	0.0136	&	-0.4513	$\pm$	0.0150	&	-1.4976	$\pm$	0.0461	&	Star-forming	\\
KPG433A	&	Sbc	&	-0.3162	$\pm$	0.0105	&	-0.4688	$\pm$	0.0581	&	-0.4791	$\pm$	0.0196	&	-1.5846	$\pm$	0.0807	&	Star-forming	\\
KPG433B	&	Sbc	&	-0.4007	$\pm$	0.0088	&	-0.4745	$\pm$	0.0271	&	-0.6150	$\pm$	0.0158	&	-1.7666	$\pm$	0.0749	&	Star-forming	\\
KPG434A	&	Sb	&	-0.1953	$\pm$	0.0084	&	-0.1419	$\pm$	0.0182	&	-0.4085	$\pm$	0.0136	&	-1.5190	$\pm$	0.0470	&	AGN	\\
KPG440B	&	Sbc	&	-0.0786	$\pm$	0.0000	&	0.5229	$\pm$	0.1657	&	-0.4296	$\pm$	0.0888	&	-1.3291	$\pm$	0.0000	&	Sy2	\\
KPG444A	&	Sbb	&	-0.3297	$\pm$	0.0254	&	-0.3689	$\pm$	0.1314	&	-0.3151	$\pm$	0.0416	&	-1.2612	$\pm$	0.1253	&	Star-forming	\\
KPG453A	&	Sb	&	-0.5125	$\pm$	0.0154	&	-0.9789	$\pm$	0.1342	&	-0.6406	$\pm$	0.0303	&	-2.6157	$\pm$	0.9549	&	Star-forming	\\
KPG455B	&	Sbb	&	0.2561	$\pm$	0.0280	&	0.6353	$\pm$	0.1704	&	0.0502	$\pm$	0.0480	&	-0.7091	$\pm$	0.0903	&	LINER	\\
KPG458A	&	Sbc	&	-0.4771	$\pm$	0.0172	&	-0.2489	$\pm$	0.0535	&	-0.4462	$\pm$	0.0266	&	-1.3659	$\pm$	0.0793	&	Star-forming	\\
KPG472A	&	Sbc	&	-0.9767	$\pm$	0.0151	&	0.4390	$\pm$	0.0128	&	-0.5645	$\pm$	0.0143	&	-1.5243	$\pm$	0.0373	&	Star-forming	\\
KPG472B	&	Sbc	&	-0.4078	$\pm$	0.0055	&	-0.5936	$\pm$	0.0120	&	-0.6977	$\pm$	0.0081	&	-2.1438	$\pm$	0.0372	&	Star-forming	\\
KPG473A	&	Sb	&	0.1025	$\pm$	0.0128	&	0.6555	$\pm$	0.0172	&	-0.0690	$\pm$	0.0206	&	-0.7704	$\pm$	0.0355	&	S-L	\\
KPG474A	&	Sb	&	-0.5142	$\pm$	0.0170	&	-0.6642	$\pm$	0.0698	&	-0.7027	$\pm$	0.0356	&	-1.9255	$\pm$	0.2235	&	Star-forming	\\
KPG477A	&	Sb	&	0.4227	$\pm$	0.0844	&	0.1440	$\pm$	0.1333	&	0.2329	$\pm$	0.1339	&	-0.3647	$\pm$	0.1966	&	LINER	\\
KPG478A	&	Sb	&	-0.2775	$\pm$	0.0192	&	-0.3219	$\pm$	0.1062	&	-0.6253	$\pm$	0.0499	&	-1.8999	$\pm$	0.3595	&	AGN	\\
KPG495A	&	Sb	&	-0.0963	$\pm$	0.0343	&	-0.1120	$\pm$	0.1161	&	-0.2400	$\pm$	0.0655	&	-1.1265	$\pm$	0.1882	&	AGN	\\
KPG496B	&	Sb	&	-0.2829	$\pm$	0.0181	&	-0.3802	$\pm$	0.0969	&	-0.4329	$\pm$	0.0346	&	-1.3222	$\pm$	0.0984	&	AGN	\\
KPG518A	&	Sb	&	-0.8578	$\pm$	0.0156	&	0.3108	$\pm$	0.0140	&	-0.4839	$\pm$	0.0150	&	-1.5081	$\pm$	0.0458	&	Star-forming	\\
KPG557A	&	Sba	&	0.3193	$\pm$	0.1119	&	0.1060	$\pm$	0.1160	&	0.3521	$\pm$	0.1453	&	-0.7773	$\pm$	0.4971	&	LINER	\\
KPG557B	&	Sb	&	-0.1795	$\pm$	0.0276	&	0.1283	$\pm$	0.0775	&	-0.3771	$\pm$	0.0615	&	-1.7788	$\pm$	0.5191	&	AGN	\\
KPG597B	&	Sbc	&	-0.3348	$\pm$	0.0244	&	-0.3667	$\pm$	0.0791	&	-0.5572	$\pm$	0.0538	&	-1.6819	$\pm$	0.2660	&	Star-forming	\\

\sidehead{Sc}
\hline
KPG021B	&	Sc	&	-0.3770	$\pm$	0.0130	&	-0.3828	$\pm$	0.0452	&	-0.5009	$\pm$	0.0222	&	-1.6011	$\pm$	0.0939	&	Star-forming	\\
KPG052A	&	Sc	&	-0.4884	$\pm$	0.0152	&	-0.8899	$\pm$	0.1115	&	-0.6032	$\pm$	0.0291	&	-1.6722	$\pm$	0.1132	&	Star-forming	\\
KPG156A	&	Sc	&	-0.4170	$\pm$	0.0086	&	-0.9228	$\pm$	0.0437	&	-0.6787	$\pm$	0.0148	&	-1.9553	$\pm$	0.0733	&	Star-forming	\\
KPG156B	&	Sbc	&	-0.4136	$\pm$	0.0665	&	-0.1190	$\pm$	0.0420	&	-0.3272	$\pm$	0.0417	&	-1.2446	$\pm$	0.0832	&	Star-forming	\\
KPG159A	&	Sc	&	-0.6847	$\pm$	0.0131	&	0.1349	$\pm$	0.0156	&	-0.3835	$\pm$	0.0148	&	-1.3963	$\pm$	0.0397	&	Star-forming	\\
KPG168B	&	Sc	&	-0.6883	$\pm$	0.0060	&	-0.0187	$\pm$	0.0075	&	-0.8003	$\pm$	0.0078	&	-2.2243	$\pm$	0.0307	&	Star-forming	\\
KPG178A	&	Sc	&	-0.3970	$\pm$	0.0604	&	-0.6943	$\pm$	0.2973	&	-0.5252	$\pm$	0.1315	&	-1.8832	$\pm$	0.5726	&	Star-forming	\\
KPG220A	&	Sc	&	-0.4434	$\pm$	0.0470	&	-0.3413	$\pm$	0.0383	&	-0.7012	$\pm$	0.1023	&	-1.8214	$\pm$	0.1622	&	Star-forming	\\
KPG230A	&	Sc	&	-0.9396	$\pm$	0.0135	&	0.3703	$\pm$	0.0124	&	-0.6510	$\pm$	0.0145	&	-1.7235	$\pm$	0.0484	&	Star-forming	\\
KPG230B	&	Sc	&	-0.4175	$\pm$	0.0411	&	-0.4150	$\pm$	0.0402	&	-0.4549	$\pm$	0.0633	&	-1.4852	$\pm$	0.0857	&	Star-forming	\\
KPG236A	&	Sc	&	-1.1876	$\pm$	0.0107	&	0.4876	$\pm$	0.0064	&	-0.8385	$\pm$	0.0108	&	-2.0797	$\pm$	0.0350	&	Star-forming	\\
KPG241A	&	Sc	&	-0.4165	$\pm$	0.0726	&	-1.4075	$\pm$	2.3641	&	-0.4338	$\pm$	0.1264	&	-1.3555	$\pm$	0.2458	&	Star-forming	\\
KPG263A	&	Sc	&	-0.1856	$\pm$	0.0000	&	-0.3118	$\pm$	0.1230	&	-0.4967	$\pm$	0.0895	&	-1.1815	$\pm$	0.0000	&	AGN	\\
KPG272A	&	Sc	&	-0.5089	$\pm$	0.0189	&	-0.2902	$\pm$	0.0607	&	-0.4501	$\pm$	0.0256	&	-1.6451	$\pm$	0.1368	&	Star-forming	\\
KPG272B	&	Sc	&	-0.4911	$\pm$	0.0085	&	-0.6236	$\pm$	0.0293	&	-0.5440	$\pm$	0.0125	&	-1.7161	$\pm$	0.0499	&	Star-forming	\\
KPG280A	&	Sc	&	-0.2986	$\pm$	0.0158	&	0.0736	$\pm$	0.0430	&	-0.3369	$\pm$	0.0252	&	-1.1528	$\pm$	0.0542	&	AGN	\\
KPG292A	&	Sc	&	-0.3666	$\pm$	0.0051	&	-0.4466	$\pm$	0.0170	&	-0.5041	$\pm$	0.0076	&	-1.6306	$\pm$	0.0226	&	Star-forming	\\
KPG293B	&	Sc	&	-0.4792	$\pm$	0.0082	&	-0.4215	$\pm$	0.0229	&	-0.5214	$\pm$	0.0126	&	-1.5532	$\pm$	0.0366	&	Star-forming	\\
KPG309A	&	Sc	&	-0.9356	$\pm$	0.0372	&	0.4175	$\pm$	0.0303	&	-0.4406	$\pm$	0.0296	&	-1.3956	$\pm$	0.0995	&	Star-forming	\\
KPG311B	&	Sc	&	-1.1629	$\pm$	0.0067	&	0.5032	$\pm$	0.0056	&	-0.9062	$\pm$	0.0075	&	-2.0387	$\pm$	0.0170	&	Star-forming	\\
KPG316A	&	Sc	&	-0.4852	$\pm$	0.0133	&	-0.4773	$\pm$	0.0349	&	-0.4391	$\pm$	0.0179	&	-1.4281	$\pm$	0.0662	&	Star-forming	\\
KPG326B	&	Sc	&	-0.8369	$\pm$	0.0162	&	0.2645	$\pm$	0.0162	&	-0.4116	$\pm$	0.0164	&	-1.3800	$\pm$	0.0402	&	Star-forming	\\
KPG332A	&	Sc	&	-0.4636	$\pm$	0.0049	&	-0.7057	$\pm$	0.0792	&	-0.5754	$\pm$	0.0290	&	-1.8044	$\pm$	0.1368	&	Star-forming	\\
KPG340A	&	Sc	&	-0.1040	$\pm$	0.0138	&	-0.2199	$\pm$	0.0327	&	-0.2439	$\pm$	0.0229	&	-1.1756	$\pm$	0.0584	&	AGN	\\
KPG352B	&	Sc	&	-0.5745	$\pm$	0.0087	&	-0.2580	$\pm$	0.0165	&	-0.4783	$\pm$	0.0111	&	-1.6564	$\pm$	0.0344	&	Star-forming	\\
KPG354B	&	Sc	&	-1.0198	$\pm$	0.0111	&	0.4307	$\pm$	0.0095	&	-0.6488	$\pm$	0.0111	&	-1.6017	$\pm$	0.0251	&	Star-forming	\\
KPG375B	&	Sc	&	-0.4157	$\pm$	0.0203	&	-0.4856	$\pm$	0.0949	&	-0.6585	$\pm$	0.0467	&	-1.6168	$\pm$	0.1686	&	Star-forming	\\
KPG396A	&	Sc	&	-0.6008	$\pm$	0.0120	&	-0.2880	$\pm$	0.0282	&	-0.4628	$\pm$	0.0151	&	-1.6624	$\pm$	0.0616	&	Star-forming	\\
KPG397A	&	Sc	&	-0.8750	$\pm$	0.0176	&	0.3096	$\pm$	0.0148	&	-0.4922	$\pm$	0.0159	&	-1.4277	$\pm$	0.0493	&	Star-forming	\\
KPG409A	&	Sc	&	-0.1843	$\pm$	0.0350	&	-0.1009	$\pm$	0.1275	&	-0.3476	$\pm$	0.0711	&	-1.7309	$\pm$	0.6074	&	AGN	\\
KPG410A	&	Sc	&	-0.2192	$\pm$	0.0430	&	-0.8206	$\pm$	0.7114	&	-0.4261	$\pm$	0.1021	&	-1.6475	$\pm$	0.6513	&	AGN	\\
KPG415B	&	Sc	&	-0.6773	$\pm$	0.0144	&	-0.0080	$\pm$	0.0260	&	-0.4111	$\pm$	0.0166	&	-1.4555	$\pm$	0.0563	&	Star-forming	\\
KPG423B	&	Sc	&	-0.3710	$\pm$	0.0097	&	-0.4824	$\pm$	0.0339	&	-0.6073	$\pm$	0.0173	&	-1.7118	$\pm$	0.0670	&	Star-forming	\\
KPG424A	&	Sc	&	-0.3510	$\pm$	0.0058	&	-0.9426	$\pm$	0.0219	&	-0.7013	$\pm$	0.0094	&	-1.9789	$\pm$	0.0352	&	Star-forming	\\
KPG440A	&	Sc	&	-0.4141	$\pm$	0.0307	&	-0.3505	$\pm$	0.1693	&	-0.2034	$\pm$	0.0349	&	-1.2652	$\pm$	0.1371	&	Star-forming	\\
KPG444B	&	Sc	&	-0.4339	$\pm$	0.0117	&	-0.4837	$\pm$	0.0294	&	-0.4150	$\pm$	0.0167	&	-1.5429	$\pm$	0.0592	&	Star-forming	\\
KPG461A	&	Sc	&	-0.9182	$\pm$	0.0144	&	0.4357	$\pm$	0.0122	&	-0.5501	$\pm$	0.0147	&	-1.5747	$\pm$	0.0404	&	Star-forming	\\
KPG511A	&	Sc	&	-0.2486	$\pm$	0.0162	&	-0.5187	$\pm$	0.1321	&	-0.3983	$\pm$	0.0318	&	-1.8039	$\pm$	0.2530	&	AGN	\\
KPG330B	&	SBc	&	-0.8000	$\pm$	0.1290	&	0.1277	$\pm$	0.0126	&	1.5051	$\pm$	0.1180	&	0.3939	$\pm$	0.1383	&	Star-forming	\\

\sidehead{Sm}
\hline
KPG212A	&	Sd	&	-1.9482	$\pm$	0.0085	&	0.8194	$\pm$	0.0041	&	-1.3579	$\pm$	0.0076	&	-2.1572	$\pm$	0.0113	&	Star-forming	\\
KPG217A	&	Sd	&	-1.2775	$\pm$	0.0541	&	0.5761	$\pm$	0.0167	&	-0.7283	$\pm$	0.0422	&	-1.7440	$\pm$	0.0463	&	Star-forming	\\
KPG249B	&	SBm	&	-0.5904	$\pm$	0.0164	&	0.2123	$\pm$	0.0144	&	-0.5536	$\pm$	0.0210	&	-1.5194	$\pm$	0.0351	&	Star-forming	\\
KPG288B	&	SBm	&	-0.3851	$\pm$	0.0048	&	-0.0627	$\pm$	0.0060	&	-0.4600	$\pm$	0.0064	&	-1.2651	$\pm$	0.0089	&	AGN	\\
KPG294B	&	SBm	&	-1.2259	$\pm$	0.0155	&	0.4646	$\pm$	0.0088	&	-0.6840	$\pm$	0.0126	&	-1.7147	$\pm$	0.0333	&	Star-forming	\\
KPG330A	&	Sd	&	-1.0493	$\pm$	0.0083	&	0.3708	$\pm$	0.0073	&	-0.9699	$\pm$	0.0106	&	-2.2974	$\pm$	0.0525	&	Star-forming	\\
KPG344A	&	Sd	&	-1.0957	$\pm$	0.0093	&	0.2902	$\pm$	0.0080	&	-0.9707	$\pm$	0.0112	&	-2.4805	$\pm$	0.0869	&	Star-forming	\\
KPG349A	&	SBm	&	-0.7375	$\pm$	0.0095	&	0.1800	$\pm$	0.0082	&	-0.5429	$\pm$	0.0109	&	-1.6739	$\pm$	0.0325	&	Star-forming	\\
KPG406B	&	Irr	&	-0.8857	$\pm$	0.0257	&	0.3590	$\pm$	0.0253	&	-0.4545	$\pm$	0.0224	&	-1.3702	$\pm$	0.0635	&	Star-forming	\\
KPG430A	&	Irr	&	-0.4817	$\pm$	0.0124	&	-0.1523	$\pm$	0.0265	&	-0.4120	$\pm$	0.0159	&	-1.5648	$\pm$	0.0612	&	Star-forming	\\
KPG438B	&	Sd	&	-0.9974	$\pm$	0.0162	&	0.4950	$\pm$	0.0129	&	-0.6597	$\pm$	0.0161	&	-1.6763	$\pm$	0.0507	&	Star-forming	\\
KPG511B	&	Sd	&	-0.5686	$\pm$	0.0141	&	-0.1076	$\pm$	0.0275	&	-0.3915	$\pm$	0.0182	&	-1.3732	$\pm$	0.0548	&	Star-forming	\\

\enddata
\tablenotetext{a}{\tiny{M.T.- Morphological Type. }}
\tablenotetext{.}{\tiny {\textbf {AGN classification denotes those
galaxies that are AGN according to the [N {\scriptsize II}] diagrams but not to the
[S {\scriptsize II}] and/or [O {\scriptsize I}].}}}  \tablenotetext{\dag}{\tiny{\textbf {L-S
classification means that galaxies fall in the separation line for
Seyfert and LINER according to [S {\scriptsize II}] and [O {\scriptsize I}] diagrams.}}}
\tablenotetext{*}{\tiny{\textbf {Type with weak broad component in
permitted lines. Seyfert quiantitative classification according to
\citet{1992MNRAS.257..677W}.}}}

\end{deluxetable}

\end{document}